# Effects of Short Scale Roughness and Wave Breaking Efficiency on Sea Spray Aerosol Production: Multisensor Field Observations[*]

Paul A. Hwang,[1] Ivan B. Savelyev,[1] Steve L. Means,[2] Magdalena D. Anguelova,[1]

Jeffrey A. Schindall,[2] Glendon M. Frick,[1] David J. Dowgiallo,[1] and Justin P. Bobak[1]

[1]Remote Sensing Division, Naval Research Laboratory, Washington DC

[2]Acoustics Division, Naval Research Laboratory, Washington DC

[1]*Corresponding author address:* Dr. Paul A. Hwang, Remote Sensing Division, Naval Research Laboratory,

4555 Overlook Avenue SW, Washington DC 20375

Email: paul.hwang@nrl.navy.mil

**Abstract**

Simultaneous measurements of sea spray aerosol (SSA), wind, wave, underwater acoustic noise, and microwave brightness temperature are obtained in the open ocean. These data are analyzed to clarify the ocean surface processes important to SSA production. Parameters are formulated to represent surface processes with characteristic length scales over a broad range, from tens of meters to a few centimeters. The result shows that the correlation coefficients between SSA properties (number, volume and flux) and surface process parameters improve toward the shortest length scale. This suggests that whereas surface wave breaking is a necessary initial and boundary condition, the final state of the atmospheric SSA properties is controlled primarily by turbulent processes characterized by the ocean surface roughness. The investigation also reveals distinct differences of the SSA properties in rising winds and falling winds, with higher efficiency of breaking production in low or falling winds. Previous studies show that the length scale of breaking waves is shorter in mixed seas than in wind seas. Combining the observations together, it is suggestive that larger air cavities are entrained in rising winds (with wind seas more likely). The larger air cavities escape before they can be fully broken down into small bubbles for the subsequent SSA production. In contrast, the shorter breakers in low or falling winds (with mixed seas more likely) trap smaller air cavities that stay underwater longer for more efficient bubble breakup by turbulence.

## 1. Introduction

The study of short-scale ocean surface roughness is of great interest to many areas of air-sea interaction and ocean remote sensing. In some applications, the important range of roughness length scale is very narrow. For example, spaceborne microwave scatterometer and radiometer are the main instruments providing global measurements of ocean surface vector winds. The microwave radar backscattering at moderate to high incidence angles are primarily contributed by the Bragg resonance mechanism (e.g., Crombie 1955; Wright 1966, 1968) and the critical length scale of the ocean surface roughness is similar to that of the radar waves: $\lambda_B = \lambda_r \left(2\sin\theta\right)^{-1}$, where $\lambda$ is wavelength, $\theta$ is incidence angle, and subscript $B$ stands for Bragg resonance and $r$ for radar. The longer surface waves also play a secondary role to backscattering through tilting the roughness patches, especially at the two ends of the incidence angle range, e.g., see Figs. 6-7 in Valenzuela (1978) and Fig. 2 in Hwang et al. (2013). This represents one extreme of the narrowness of critical roughness length scale. For instance, at 45° incidence angle, the L-, C-, X- and Ku-band (~1.4, 5.5, 10, 14 GHz) Bragg wavelengths are 0.15, 0.039, 0.021 and 0.015 m. Their response to wind forcing and modification by other environmental parameters such as longer scale waves and currents are different, thus we can expect different degrees of sensitivity in the retrieval of geophysical parameters such as wind speed, surface current and wave properties using radar backscattering of different frequency bands.

For some processes, the range of important roughness length scale is broader. For example, in the passive microwave radiometer measurement of ocean surface emission, the contributing bandwidth of surface waves extends to a factor of about 3~10 times shorter and longer than the electromagnetic (EM) wavelength depending on the incidence angle, e.g., see Figs. 1-3 of Johnson and Zhang (1999).

Similar to remote sensing processes, in air-sea interaction, several studies have shown that the correlation of gas transfer velocity with ocean surface mean square slope (MSS) integrated over some range of wavenumber $k$ is considerably better than its correlation with wind speed (e.g., Bock et al. 1999; Frew et al. 2004, 2007). Frew et al. (2004) quantify the difference with field data: referenced to the wind-speed correlation coefficient of about 0.77, the MSS correlation coefficient improves from 0.81 ($k$ range 40-100 rad m$^{-1}$) to 0.92 ($k$ range 400-800 rad m$^{-1}$).

In a recent paper, Savelyev et al. (2014), referred to as S14 in the following, present an analysis of SSA measurement onboard the Floating Instrument Platform (FLIP) conducted by the Naval Research Laboratory (Breaking Wave Experiment: BREWEX). Among the reported results is the total sea spray source flux (SSSF) $\Sigma F$ integrated over the range of dry radius between 0.5 and 12 μm. Referenced to the wind speed, the $\Sigma F$ seems to separate into two or more distinct populations. A significant improvement in bringing the multiple populations together is achieved when the $\Sigma F$ is plotted against the roughness component of the microwave (10.7 GHz) brightness temperature $\Delta T_b$. They hypothesize that the observation may suggest that the processes important to SSA generation are probably better characterized by the short scale ocean surface roughness, although the exact physical mechanisms remain to be identified.

In this paper, we continue exploring the length scale issue of ocean processes important to SSA production. In addition to wind and microwave brightness temperature measurements, the BREWEX data suite includes surface waves obtained by a pressure sensor at 3 to 5 m depth and underwater acoustic noise in the frequency range between 300 and 2400 Hz. The acoustic system is a 32-element vertical hydrophone array with an 18.75-m aperture and the top element is at 9 m below the mean water surface. Unfortunately, the low frequency portion -- less than 1200 Hz -- is contaminated by the noise from the FLIP so only the high frequency portion is used in this study.

The dynamic pressure of wave motion decays exponentially away from the interface, therefore the useful surface wave signal is limited to wavelengths longer than about twice the sensor depth (6 to 10 m minimum wavelength in the present case). Wave-induced acoustic noise is likely produced at the instant of bubble detachment or splitting ("screaming infant microbubbles") (e.g., Pumphrey 1989; Prosperetti et al. 1989; Medwin and Beaky 1989; Pumphrey and Elmore 1990; Longue-Higgins 1992; Medwin and Clay 1998). The underwater acoustic noise measurement therefore captures the bubble-entraining wave breaking, with dominant length scale of probably a few meters, as estimated from the first moment of the breaking front statistics characterizing the ocean surface whitecap coverage (Phillips 1985) and the velocity and length scales derived from analyses of radar sea spikes (Frasier et al. 1998; Hwang 2007; Hwang et al. 2008a, b).

Considering wind as the common driving force, the observations of microwave brightness temperature, acoustic noise and surface waves thus serve as proxies of processes responding to surface roughness of three distinctive ranges of length scale, respectively, a few centimeters, a few meters and tens of meters and longer. For brevity, we denote the three ranges as O(0.01 m), O(1 m) and O(10 m) in the remainder of this paper. Our analysis shows that the correlation coefficients between SSA properties and the process proxies improve toward the shortest length scale. Also, by sorting the data into rising wind, falling wind, and quasi-steady categories, there is a clear episodic behavior of the SSA dependence on the windsea energy dissipation rate $E_t$. The efficiency of SSA production by wave breaking is much better in low or falling winds than in rising winds. A plausible explanation of these two primary results (short scale roughness and breaking efficiency) has been summarized in the abstract and the analyses leading to the explanation are described in this paper.

In the following, Sec. 2 gives a brief recapitulation of the experiment; a detailed description has been presented in S14. Sec. 3 describes the results of the suite of measurements: wind, wave, SSA, brightness temperature and underwater acoustic noise. These measurements are then combined to explore the mechanisms of SSA generation and distribution. Sec. 4 discusses the issues of roughness scales, breaking efficiency and the implication on air-sea interaction processes in general and SSA generation in particular, and Sec. 5 is a summary.

## 2. Experiment

The Breaking Wave Experiment (BREWEX) is conducted onboard the free-drifting FLIP in the Pacific Ocean about 200 km west of California coast from 17 April to 3 May 2012 (Day 108 to 124, 1 Jan being Day 1). The overall goal of the experiment is to conduct a variety of collocated measurements aimed at identifying specific signatures of active and residual phases of oceanic whitecaps utilizing visible, infrared, microwave, as well as acoustic sensing. The first results of the BREWEX data analysis on SSA and passive microwave emissions from the ocean surface are presented in S14. The analysis focuses on the later period when the wind turned to be from north and stayed northerly for the next four days (Day 118 to 122) and the SSA is not contaminated by the land or surf sources.

Figure 1 depicts the trajectory of the FLIP, highlighted in dark color is the duration of the northerly wind. Also shown in the figure are the locations of four National Data Buoy Center (NDBC) buoys used to compare with the wind and wave measurements onboard FLIP. The depth of the pressure sensor for surface wave measurements is initially at 3 m but lowered to 4 then 5 m in the later part of the experiment as wind and wave increase. The pressure signal is sampled at 10 Hz but the exponential decay of the dynamic wave-induced pressure signal places the upper limit of the useful frequency range at about 0.4 Hz, similar to that of the NDBC buoys. As illustrated in Fig. 2, the measurements of wind speed and wave properties (significant wave height $H_s$ and spectral peak wave period $T_p$) are in general agreement with the NDBC buoys. A motion sensor package is installed to measure the six degrees of the FLIP motion. The analysis of several data segments during high sea state shows that the wave spectral correction is on the order of a few percent at most, consistent with earlier analyses of the FLIP motions. For example, Smith and Rieder (1997) conclude that "... The resultant wind stress contribution is … more than two orders of magnitude smaller than the measured wind stresses.

Indeed, FLIP **is** a stable platform." The motion correction is therefore not performed in our wind and wave measurements.

For the FLIP data, the significant wave height $H_s$ is computed as four times the square root of the elevation variance from integrating the wave spectrum, which is calculated from the pressure signal with the exponential decay correction. The peak wave period $T_p$ is obtained using Young's (1995) algorithm:

$$\omega_p = \frac{\int_0^\infty \omega S^4(\omega)\,d\omega}{\int_0^\infty S^4(\omega)\,d\omega}, \tag{1}$$

where $\omega$ is angular frequency $2\pi T^{-1}$, $T$ is wave period, $S$ is spectrum, and subscript $p$ indicates the value at spectral peak. The peak period obtained from the spectral integration method is less noisy compared to simply picking the maximum component of the spectrum or several other designs of peak frequency computation; the detail is discussed in Young (1995). Although there are differences in the fine details between the FLIP and NDBC time series because of the spatial separation (several hundred km apart), the large scale features are similar in all measurements.

The durations of FLIP wind and wave recordings are (time) $t$ = 113.02-121.81 d and 114.90-121.67 d, respectively. In Fig. 2, the 10-min average results are merged together and interpolated to the same reference time of the latter period. The buoy data are reported hourly with wave spectrum computed from 20 min data and wind averaged over 8 min. As discussed in Sec. 1, the wind turned to be from the north on Day 117. For the remaining period of the experiment, the wind direction remains steady and $U_{10}$ varies from 1.9 to 15.6 m s$^{-1}$ with some fluctuation. The wave condition is expected to change from swell-dominant in low winds to windsea-dominant in high winds. The significant wave height $H_s$ exceeds 4 m during the high wind period, the minimum $H_s$ is 0.98 m (Fig. 2c).

## 3. Results

### 3.1. Wind-wave energy dissipation

To emphasize the relevance to wave breaking, the combination of wind and wave data is represented by the windsea energy dissipation rate (Hwang and Sletten 2008; Hwang 2009). The separation of windsea and swell components is based on the spectrum integration method (Hwang et al. 2012). The result of applying the method to the FLIP data is shown on the left column of Fig. 3: (a) displays the time-frequency spectral plot; (b) the significant wave heights of the windsea and swell components; and (c) the peak wave periods of the windsea and swell components. There is a gap (the dark band in the spectral plot of Fig. 3a) in the wave data between $t = 118.043$ and $118.143$ d when the pressure sensor depth is adjusted from 3 m to 4 m to prevent breaching. Another adjustment to 5 m is done near $t = 119$ d without stopping data recording. The time series of wind speed and windsea peak frequency $f_{pw}$ obtained from swell-sea separation are superimposed on the wave spectrum plot (Fig. 3a) to show the complex evolution of surface wave spectrum in response to fluctuating wind forcing.

The windsea component of the wave spectrum is used to estimate the spectrally integrated energy dissipation rate attributed to wave breaking (Hwang and Sletten 2008):

$$E_t = \alpha \rho_a U_{10}^3, \text{ with } \alpha = 0.20 \omega_{\#}^{3.3} \eta_{\#}, \tag{2}$$

where $\rho_a$ is air density (about 1.22 kg m$^{-3}$), $\omega_{\#} = \omega_p U_{10} g^{-1}$ is the dimensionless peak frequency, $\eta_{\#} = \eta_{rms}^2 g^2 U_{10}^{-4}$ is the dimensionless windsea elevation variance, $U_{10}$ is reference wind speed, $\omega_p$ is the windsea spectral peak frequency, $\eta_{rms}$ is the root mean squares (rms) surface elevation of the wind sea spectrum, and $g$ is the gravitational acceleration.

The estimated wave energy dissipation rate is shown in Fig. 3d. The results displays the expected feature of low energy dissipation in swell-dominant or low wind periods (e.g., $t < 118$ d) and high dissipation in younger sea and high winds (these two factors are closely correlated and will be further discussed in the next paragraph). Note that the total wave height $H_s$ includes windsea and swell components and may not always be a good indicator of energy dissipation. For example, the trends of $H_s$ and $E_t$ in the periods 117.5-118 d and 120+ d are noticeably different (Fig. 3d). In general, in a sustained increasing wind event, $H_s$ and $E_t$ grow steadily following the same trend. Passing the peak wind speed and in the falling wind phase of the event, $H_s$ continues to grow for a few hours due to nonlinear energy transfer but $E_t$, being the property of the active wind-generated waves, would decay at the pace of $U_{10}$ (Figs. 3d and 3e).

For our deep water wave condition, the dimensionless frequency $\omega_\#$ is also the inverse wave age $U_{10}c_p^{-1}$ (a younger sea corresponds to larger value in $\omega_\#$). For the period to be focused in this study ($t$>117.5 d with northerly winds and most sensors online), $\omega_\#$ is less than 1.8. Higher wind generally produces younger sea as expected from the slower variation of waves compared to winds – thus $U_{10}c_p^{-1}$ reflects more the $U_{10}$ fluctuation than the $c_p$ fluctuation (Fig. 3e). This is especially true in the situation of low or falling winds: for the latter, waves continue to grow even after wind passed its peak; for the former, the windsea spectrum is hidden under the strong swell and cannot be separated.

The magnitude of $E_t$ is determined mainly by the cubic wind speed relationship (2); the $\alpha$ value varies only slightly in field conditions: $\alpha \approx \left(4.7 \pm 1\right) \times 10^{-4} \rho_a \approx \left(5.6 \pm 1.2\right) \times 10^{-4}$ (Hwang and Sletten 2008). For low winds and windsea failed to be separated, the resulting $E_t$ estimate may bias high or low depending on the mixed wave conditions. Based on our experience, the $E_t$ calculation of field data for $U_{10}$ < 5 m s$^{-1}$ is generally impacted by the difficulty in separating windsea and swell, especially in swell dominant conditions. Using the approximate $\alpha$ formula above, the corresponding $E_t$ is less than $(7.0 \pm 1.5) \times 10^{-2}$ W m$^{-2}$ for $U_{10} < 5$ m s$^{-1}$. Previous analyses of field data show a sharp $E_t$ dropoff at $U_{10}$ less than about 5 to 7 m s$^{-1}$ (Hwang and Sletten 2008; Hwang 2009).

### 3.2. Brightness temperature

Microwave emission from the ocean surface is modified by the roughness and white foam produced by breaking waves (e.g., Hollinger 1971; Smith 1988). Because both ocean surface roughness and wave breaking are closely associated with wind speed, it has been used for ocean surface wind retrieval since the 1970's and an extensive body of literature exists (e.g., Wentz 1975; Ulaby et al. 1981; Wentz et al. 1986; Yueh 1994a, b, 1995; Johnson and Zhang 1999; Johnson 2006). A short discussion has been given in S14. Here we only show the processing result. The frequency of the microwave radiometer is 10.7 GHz operated in both horizontal and vertical polarizations. It is placed at about mid-point of the FLIP's port-side 18-m boom and points at 45° toward the ocean surface.

For geophysical applications, the surface emissivity is frequently represented by the brightness temperature $T_b$, which is the product of emissivity $e_p$ and surface temperature $T_s$: $T_{bp} = e_p T_s$, where subscript $p$ is polarization ($H$ or $V$ for horizontal or vertical). The measured $T_{bp}$ is composed of the flat surface emission $T_{bp0}$ and a small correction $\delta T_{bp}$ contributed by wind and wave disturbances of the flat sea surface. There have been many algorithms established to compute the emissivity of unperturbed fresh or sea water (e.g., Klein and Swift 1977; Meissner and Wentz 2004). For the typical salinity (35 psu) and surface temperature (13° C) in our experiment, the two algorithms mentioned above produce a slight difference of the flat surface ($T_{bH0}$, $T_{bV0}$): (81.4K, 139.7K) for the former and (81.8K, 140.2K) for the latter. In the remainder of this paper, the algorithm of Meissner and Wentz (2004) is used to remove the flat surface emission from the measured $T_{bp}$. The separation of foam and roughness components of the brightness temperature measurements uses the equivalent medium method described by Hwang (2012). A summary is given in Appendix A.

Figure 4a shows all the $\delta T_{bp}$ data obtained in this experiment, each data point represents 20-min average. The variation of $\delta T_{bp}$ generally tracks $U_{10}$ very well and the trend is in general agreement with theoretical predictions (see also Fig. 9 in S14). The analytical result shown in Fig. 4b is based on the SPM/SSA (small perturbation method, small slope approximation) model (e.g., Yueh 1994a, b, 1995; Johnson and Zhang 1999; Reul and Chapron 2001) and separates the roughness and foam components (Hwang 2012; Appendix A). There is some apparent offset between data and model, most likely caused by reflection of the atmospheric downwelling, which is not corrected due to the lack of information such as detailed cloud cover, type and atmospheric constituents. Another source of error is the usage of constant sea surface temperature (13° C) and salinity (35 psu) for computing the flat surface value of the brightness temperature, due to the lack of continuous measurements. The magnitude of the offset is similar to those reported in other field experiments (e.g., Hollinger 1977; Swift 1974; Sasaki et al. 1987; Camps et al. 2004).

For the range of wind speed encountered in the $T_b$ measurements (less than 15 m s$^{-1}$), the foam contribution is relatively minor and weakly dependent on polarization. The difference between $\delta T_{bH}$ and $\delta T_{bV}$, represented by $\Delta T_b$, is dominated by the roughness effect (Fig. 4c). A fringe benefit from this difference operation is that the uncorrected atmospheric effects and errors of flat surface emissivity computation are mostly removed because they are only weakly sensitive to polarization. The agreement between modeled and measured $\Delta T_b$ is very good.

### 3.3. Underwater acoustic noise

The close association of underwater acoustic noise and surface wind and wave conditions has been recognized for many decades (e.g., Knudsen et al. 1948; Wenz 1962). The extensive collection of field observations have established the general rule of "5 dB per octave": the noise level $P$ decreases 5 dB as frequency $f$ doubles and increases 5 dB as wind speed doubles. The general rule can be written as $P(f, U_{10}) \sim f^{-1.7} U_{10}^{1.7}$. While a good rule of thumb, it is also recognized that the underwater acoustic noise level is dependent on location, season, day or night, biological activities, shipping, and many other correlated or uncorrelated factors (e.g., Clay and Medwin 1977; Urick 1983, 1984; Medwin and Clay 1998).

One of the primary mechanisms of the wind-related noise is identified to be the ringing (shock excitation) of bubble formation at the time of detaching from the entrained air cavity (Pumphrey 1989): "screaming, infant microbubble" as described by Medwin and Clay (1998, pp. 15, 334). The high efficiency and precise resonance of underwater bubble oscillations have been fascinating and actively-researched subjects of underwater acoustics and breaking wave studies (e.g., Minnaert 1933; Clay and Medwin 1977; Kerman 1984, 1988, 1993; Medwin and Beaky 1988; Prosperetti 1988; Pumphrey 1989; Medwin and Daniel 1990; Longuet-Higgins 1991, 1992, 1993; Crum 1995; Lamarre and Melville, 1994a, b; Buckingham and Potter 1995; Dahl and Jessup 1995; Felizardo and Melville 1995; Melville 1996; Medwin and Clay 1998; Deane and Stokes 2002, 2006, 2010; Zhao et al. 2014; and references therein).

In this study, the acoustic noise data is obtained by a vertical hydrophone array with 32 elements configured into three center-nested apertures. The element spacing for the three apertures are 1.25, 0.625, and 0.3125 m and yield design frequencies of 600, 1200, and 2400 Hz, respectively. The center of each aperture is nominally 18.2 m below the surface.

The raw data of each element is sampled at 12.5 kHz. Each channel is Fast-Fourier transformed (FFT) yielding time and frequency resolutions of 0.1 s and 10 Hz, respectively. To reduce noise from directions other than the overhead surface, the array data is phase-processed to listen to noise from the end-fire beam (Urick 1983, pp. 54-58).

The array is deployed vertically between the face and port booms with its axis roughly 13 m from the hull of the FLIP. Due to the width of the end-fire beam as a function of range, lower frequency noise generated at the FLIP hull protrudes into the end-fire beam of the array. Thus, the data reported in this paper is limited to the frequency range between 1250 and 2350 Hz. The acoustic energy received is normalized by the frequency-dependent cross-sectional area of the surface spanned by the end-fire beam (circles of radii ranging between 13 m and 9 m for the frequencies of interest here) and averaged over 100-Hz frequency bands. The modulation of the acoustic reception surface area by the ocean waves passing over the acoustic array is found to be insignificant even for the most severe sea state period (Appendix B).

Figure 5a shows the noise power spectra sorted by wind speed. The coefficients $A_f$ and $a_f$ of least-squares fitted power-law function: $P(f;U_{10}) = A_f f^{a_f}$ are shown in Figs. 5b and 5c, respectively. The frequency range in our data is at the boundary between shipping (10 to 1000 Hz) and wind wave action (50 Hz to 20 kHz) and field data generally show a local peak or flattening of the spectral shape, e.g., see the large collection discussed in Wenz (1962) and recent results of undersea noise in hurricane conditions (Zhao et al. 2014). Experiments of acoustic noise produced by breaking waves conducted in anechoic wave flumes (e.g., Medwin and Beaky 1989; Crum 1995) also show the leveling off of the noise spectrum near 1000 Hz. The frequency exponent $a_f$ in our data set is close to 0 at low wind speed and gradually reduces to about -0.7 and -0.5 in mid to high winds.

Figure 6a shows the wind speed dependence of the noise power spectral components. In the low wind region ($U_{10}<\sim5$ m s$^{-1}$), the wind speed trend is weak, indicating that the wave-induced noise signal is barely above the ambient background. For $U_{10} > \sim 5$ m s$^{-1}$, the steady trend of increasing with wind is clear. The coefficients $A_U$ and $a_U$ of least-squares fitted power-law function: $P\left(U_{10}; f\right) = A_U U_{10}^{a_U}$ are shown in Figs. 6b and 6c, respectively. The fitting is performed for cases with $U_{10} \leq 5$ m s$^{-1}$, $U_{10} \geq 5$ m s$^{-1}$ and all wind speeds. The wind speed dependence is close to $P \sim U_{10}^{1.7}$ except for the low-wind group.

### 3.4. Aerosol properties

The study of sea spray aerosol or sea salt aerosol (SSA) is of importance to many research areas such as air-sea mass, heat and momentum exchanges, cloud and weather, climate, and atmospheric optical properties. An extensive survey of the subject by Lewis and Schwartz (2004) includes more than 1800 references; the more recent reviews by O'Dowd and de Leeuw (2007) and de Leeuw et al. (2011) feature 65 and 133 references, respectively. The extensive collective research has shown that uncertainty of a factor of 4 to 5 is common in measuring and parameterizing the production flux of SSA over a wide range of aerosol sizes.

Although wind speed is generally used as the starting parameterization factor, it is also recognized that wind speed alone is inadequate to account for variations of sea state, air-sea boundary layer stability, water temperature, surfactants, …, among many other factors that cause quantitative changes of the SSA properties. There are many alternative parameterization functions based on, e.g., combined wind and wave parameters, whitecaps, or bubble size spectrum (e.g., see the reviews by Hoppel et al. 2002; Lewis and Schwartz 2004; de Leeuw et al. 2011; Ovadnevaite et al. 2014). S14 presents a parameterization function based on the microwave radiometer brightness temperature, and suggests using such an approach for global SSA monitoring. Here we revisit the aerosol data collected during BREWEX. In addition to the aerosol flux given in S14, we also present results of number and volume size distributions as well as a different calculation of the SSA flux.

The aerosol size spectra are obtained with a forward scattering spectrometer mounted on the starboard boom of the FLIP. The location is about 7 m away from the FLIP hull and 7.3 m above the mean water level. The detail of the measurement and data processing has been described in S14 and only a brief summary is given here.

The optical scattering spectrometer measures the SSA spectra in 4 size ranges: (in radius $r$) 1.5-23.5, 1.0-16.0, 0.5-8.0, and 0.25-4.0 μm. Combining the four size ranges, S14 presents the computed sea spray source function in the dry radius $r_{dry}$ range of 0.51 to 12.14 μm (with 0 registered in the two largest size bins – 11.09 and 12.14 μm). Conversion of the measured (*in situ*) particle radius $r_{is}$ to the dry radius with the relative humidity (RH) input uses the algorithm of Gerber (1985), prepared as a lookup table of triplets [ $r_{is}$ , RH, $r_{dry}$ ] for each size bin of the optical spectrometer; the RH in the table ranges from 0 to 100% in 1% increments.

The aerosol flux is computed by the product of aerosol number $N(r)$ and its settling velocity $v_g(r)$ evaluated with the measured (*in situ*) $r_{is}$ at the ambient relative humidity:

$$F_{is}(r) = N(r) v_g(r_{is}).$$ (3)

The settling velocity is calculated with the approximation equation

$$v_g(r) = 0.01 \left( \frac{r}{8.5} \right)^2,$$ (4)

with $v_g$ in m s$^{-1}$ and $r$ in μm (Lewis and Schwartz 2004, p. 66).

We also examine a different formulation of the SSSF:

$$F_U(r) = N(r) v_d(r_{80}).$$ (5)

The dry deposition velocity $v_d$ considers the turbulence factor and for the coarse aerosols in this study, it is approximated by

$$v_d\left(r_{80}\right) = 0.01\left[v_g\left(r_{80}\right) + \frac{U_{10}}{4.6}\right],$$ 
(6)

where $v_d$ is in m s$^{-1}$, $r$ is in μm, $U_{10}$ is in m s$^{-1}$ and $v_g$ is evaluated at $r_{80}$: the equivalent radius at 80% RH (Lewis and Schwartz 2004, p. 283). Extensive discussions on the various formulas for $v_d$ have been published (e.g., Sehmel 1980; Slinn and Slinn 1980; Smith et al. 1993; Hoppel et al. 2002; Lewis and Schwartz 2004) and they will not be repeated here. The approximation $r_{80} = 2r_{dry}$ (Lewis and Schwartz 2004, pp. 53-54) is used for (5) and (6). As described in Appendix C, some small difference is found between the humidity correction formula of Gerber (1985) and the approximation formula of Lewis and Schwartz (2004). The small difference is not expected to change the conclusions of our analyses.

Unless stated otherwise, the results are presented in terms of $r_{80}$ in this paper. The subscripts in the two different estimates of SSA flux, given as $F_{is}$ and $F_U$, may be dropped when distinction between the two is not needed. The primary difference in these two different computations of $F(r)$ is in the small size range, of which the settling velocity is small and the turbulent component of the deposition velocity (proportional to $U_{10}$) raises considerably the magnitude of $F_U$ compared to that of $F_{is}$.

Figures 7 show the SSA volume $V$, number $N$ and flux $F$ based on the measurements of the first size-range of the optical scattering spectrometer (1.5-23.5 μm, which results in the $r_{80}$ range of about 1.3 to 20 μm) for several average wind speeds from 4 to 15 m s$^{-1}$, the wind speed bin size used in the computation is ±1 m s$^{-1}$. The SSA size spectrum shows a general and monotonic increase with wind speed in all the size components in our data with an exception of the lowest wind speed bin, of which the magnitude of the large particles appears to be abnormally high. This is probably caused by the higher RH in some of the data segments in this low wind speed bin: the average RH is 87% with a maximum of 96% in some segments. Condensation on the sensor intake area causing measurement contamination may have occurred under such condition of high humidity and low wind speed. Further discussion of the humidity influence is deferred to Sec. 4.1.d.

The SSA dependence on $U_{10}$ follows the power-law relationship very well, whether in terms of volume, number or flux (Fig. 8). The relationship, in fact, can be used to construct empirical model functions (EMFs) for the SSA size spectra:

$$X\left(r, U_{10}\right) = A_X\left(r\right) U_{10}^{a_X(r)}, \tag{7}$$

where $X$ can be $dV / d\ln r$, $dN / d\ln r$ or $dF / d\ln r$. Least-squares fitted $A_X$ and $a_X$ are listed in Table 1, and they can be expressed as polynomial functions (Fig. 9):

$$Y\left(Z\right) = p_{Z1}Z^3 + p_{Z2}Z^2 + p_{Z3}Z + p_{Z4}, \tag{8}$$

where $Z$ is $A_X$ or $a_X$, and $P_{Z1}$ to $P_{Z4}$ are the fitting coefficients (Table 2). The wind speed dependence in terms of the exponent $a$ of the power-law function (7) ranges from about 1 for $dV / d\ln r$, $dN / d\ln r$ or $dF_{is} / d\ln r$ and 2 for $dF_U / d\ln r$ at $r$=~2 μm to between 4 and 5 at $r$=~14 μm, and shows a clear trend of increasing nonlinearity with the aerosol size (Fig. 9b).

Equations (7), (8) and Table 2 complete the EMFs of the SSA volume, number and flux size spectra; constructed with data in the radius range of 2 to 14 μm and wind speeds between 5 and 15 m s$^{-1}$. (Cases with $U_{10}$<5 m s$^{-1}$ are not used in data fitting because of their high noise. Also both ends of the aerosol size range are excluded because after humidity adjustment, the uniform $r_{is}$ size range of 1.5 to 23.5 μm becomes irregular in the $r_{80}$ range).

Figure 10 shows the comparison between our $F_U$ EMF computed for $U_{10}$=5, 10 and 15 m s$^{-1}$ and measurements processed in 5±1, 10±1 and 15±1 m s$^{-1}$ $U_{10}$ bins. The local bump in the large size region of the lowest wind speed data is not observed in higher wind speeds. The exact cause is not certain but may be related to the high relative humidity and low wind conditions, as commented earlier in the discussion of Fig. 7.

Also shown in the figure are the results based on the models of Monahan et al. (1986): M86, Smith et al. (1993): S93; and Lewis and Schwartz (2004, eq. 2.1.9 and p. 341): L04, respectively

$$\frac{dF_{M86}}{d\ln r} = 1.373 U_{10}^{3.41} r^{-2} \left(1 + 0.057 r^{1.05}\right) 10^{1.19\exp\left\{-\left[\frac{(0.38-\log r)}{0.65}\right]^2\right\}}, \tag{9}$$

$$\frac{dF_{S93}}{d\ln r} = \frac{1}{\ln 10}\left\{\left[1400\exp\left(0.16 U_{10}\right)\exp\left(-3.1\left(\ln\frac{r}{2.5}\right)^2\right)\right] + \left[0.76\exp\left(2.2 U_{10}^{0.5}\right)\exp\left(-3.3\left(\ln\frac{r}{11}\right)^2\right)\right]\right\}, \tag{10}$$

and

$$\frac{dF_{L04}}{d\ln r} = \frac{1.54\times10^4}{\ln 10}\left(\frac{U_{10}}{10}\right)^{2.5}\exp\left\{-0.5\left[\frac{\ln\left(r/0.3\right)}{\ln 4}\right]^2\right\}. \tag{11}$$

Our data registered much smaller spectral magnitude in the larger size bins compared to those modeled by M86, S93 and L04. For the smaller size region in our measurements, the size spectral magnitude is significantly impacted by the choice of deposition velocity formulas (compare Figs. 7c and 7d). The overall agreement between our data and the modeled results of M86, S93 and L04 is comparable to the agreement between different models.

The more complete formulation of the deposition velocity used in this paper compared to the one used in S14 leads to a better agreement in the SSA flux estimates of the dry deposition method and the vertical gradient method. In S14, the dry deposition estimate $F_{is}$ is reported to be lower than that obtained by the vertical gradient method by a factor of about 10 to 30 depending on the aerosol size. The vertical gradient method is described in S14 and used to estimate $dF/d\ln r$ from vertical concentration profiles for a time period of several hours with steady wind speed of $U_{10} = 11$ m s$^{-1}$ (see their Fig. 6). Those same points are shown in our Fig. 10 with asterisks. They appear to be in much closer agreement with our estimates of the SSA flux $F_U$. This agreement is an additional validation of both methods, and also suggests that the dry deposition calculation $F_U$ is more accurate than $F_{is}$.

### 3.5. Combining multi-sensor measurements

The simultaneous time series of 20-min average underwater acoustic noise, SSA flux, windsea energy dissipation rate, roughness component of microwave brightness temperature, and wind speed are displayed together in Fig. 11, in linear scale on the left column and logarithmic scale on the right column.

Figure 11a shows two representations of underwater acoustic noise: the spectral levels at 2350 Hz and the average over the frequency range between 1250 and 2350 Hz, respectively $P_{2350}$ and $P_{<f>}$. The magnitude of $P_{<f>}$ is slightly higher than $P_{2350}$. Fig. 11b displays two estimates of the integrated SSA flux: $\Sigma F_U$ and $\Sigma F_{is}$; these integrated fluxes represent the atmospheric aerosol loading in the $r_{is}$ range of 1.5 to 23.5 μm. The magnitude of $\Sigma F_{is}$ is about 20 times smaller than that of $\Sigma F_U$. The $E_t$ and $\Delta T_b$ are given in Figs. 11c and d, respectively. They all follow similar trends tracking the $U_{10}$ fluctuation (Fig. 11e). Vertical reference lines are drawn in Fig. 11f marking six wind fluctuation events (rising, falling, quasi-steady) for comparison of various measurements. The analysis of the six events will be further detailed later. From this point on, unless stated otherwise, the underwater acoustic noise is represented by $P_{<f>}$, the SSA flux is represented by $\Sigma F_U$ and subscript $U$ may be dropped. We also show the relative humidity RH time series in Fig. 11e. As will be seen later, some of the peculiar behavior of the SSA results coincides with the RH variation.

The relationships between $E_t$, $P_{<f>}$, $\Sigma F$, $\Delta T_b$ and $U_{10}$ can be represented by power-law functions as shown in Fig. 12a. For reference, line segments with $U_{10}^2, U_{10}^3$ and $U_{10}^5$ are superimposed; they also serve as partitions separating the various measurements. We are especially interested in the $E_t$ parameter in this paper as will be further explained in the next section. Also shown in the figure is the set of curves corresponding to the empirical relationship $E_t = \left(4.7 \pm 1\right) \times 10^{-4} \rho_a U_{10}^3$ derived from examining several field datasets as discussed in Hwang and Sletten (2008) and Hwang (2009). The $E_t$ dependence on wind speed is also close to cubic in the present measurements in high winds.

The large scatter of $\Sigma F\left(U_{10}\right)$ is rather unsettling (Fig. 12a). A closer inspection reveals that the data is constituted of several populations following similar wind speed dependence but with different absolute magnitudes. In Fig. 12b, the temporal evolutions of $P_{<f>}$, $\Sigma F$, $E_t$ and $\Delta T_b$ in several interesting time segments are highlighted by different colors (increasing winds, decreasing winds, quasi-steady; as marked by vertical line segments in Fig. 11). Comparing the results of the consecutive rising and falling wind events of 118.1-119.25 d and 119.25-119.83 d, a "hysteresis" in $\Sigma F\left(U_{10}\right)$ is quite obvious. A similar response lag is found in the acoustic noise and roughness component of brightness temperature, but conspicuously mixed in the windsea energy dissipation rate. In the high wind region in particular, the energy dissipation rate is generally less in falling winds than in rising winds for the same wind speed. This is reasonable considering that rising winds carry additional forcing of wind speed acceleration. But then why the reverse trend in SSA flux, acoustic noise and surface roughness? This will be further explored (Sec. 4) after investigating the correlation coefficients between SSA flux and several forcing parameters.

Figure 13 shows the scatter plots of $\Sigma F$ vs. $U_{10}$, $E_t$, $P_{<f>}$, and $\Delta T_b$; the correlation coefficients $R^2$ are 0.72, 0.57, 0.82 and 0.90, respectively. In the following, $R^2_{xy}$ represents the correlation coefficient between variables $x$ and $y$, and shorthand $F$ for $\Sigma F_U$, $U$ for $U_{10}$, $P$ for $P_{<f>}$, and $T$ for $\Delta T_b$, thus $R^2_{FU}$, $R^2_{FP}$, ... and so on. In this figure, we use only the data segments with all measurements ($\Sigma F$, $U_{10}$, $E_t$, $P_{<f>}$, and $\Delta T_b$) available simultaneously: the most constraining data series is $\Delta T_b$ (Fig. 11). The correlation is clearly much better between $\Sigma F$ and $\Delta T_b$ than that between $\Sigma F$ and $U_{10}$, $E_t$ or $P_{<f>}$. The improvement in correlation from $U_{10}$ to $\Delta T_b$ is also observed in other expressions of the SSA properties such as the integrated number $\Sigma N$ or volume $\Sigma V$. Fig. 14 shows $\Sigma N(U_{10})$ and $\Sigma V(U_{10})$ on the top row, and $\Sigma N(\Delta T_b)$ and $\Sigma V(\Delta T_b)$ on the bottom row, illustrating the consistent increase of the correlation coefficient from $U_{10}$ to $\Delta T_b$ parameterization of the SSA properties. The correlation between the SSA properties with $E_t$ is worse than those with $U_{10}$, and the correlation between the SSA properties and $P_{<f>}$ is better than those with $U_{10}$ but worse than those with $\Delta T_b$ (Fig. 15), the trend is the same as that illustrated in Fig. 13.

The result of the correlation analysis is rather surprising considering that $E_t$ and $P_{<f>}$ (as well as $U_{10}$) are closely connected with wave breaking, and that bubble bursting from wave breaking has been considered an important (or maybe the most important) source of SSA (e.g., Monahan et al. 1986; Lewis and Schwartz 2004; de Leeuw et al. 2011). The implication seems to be that although wave breaking is a necessary initial and boundary condition for getting the SSA into air, the turbulent transport processes controlled by small scale surface roughness are more important in determining the final state of the SSA in the atmospheric boundary layer.

However, before settling on that conclusion, we need to examine the experimental conditions in more detail. In particular, the driving wind is far from the ideal homogeneous and steady state generally assumed in theoretical works, but rather a combination of typical natural fluctuations with rising, falling and quasi-steady wind events of various durations (Fig. 11f). These are highlighted with different colors or symbols in Fig. 12b, as well as in Figs. 13 to 15. There is also the fluctuation of relative humidity (Fig. 11e) that seems to coincide with some of the observed anomalies, in particular, the abnormally low values of $\Sigma F$, $\Sigma N$, and $\Sigma V$ during the period 118.1-119.23 d (Figs. 13-15) when the RH is the lowest (Fig. 11e). We explore further the effects of these $U_{10}$ and RH variations on the result of the correlation analysis next.

## 4. Discussion

### 4.1. Efficiency of wave breaking and air-sea interaction

The result from the SSA correlation analysis (Figs. 13-15) is quite puzzling at first glance. Specifically, when compared to the $\Sigma F(U_{10})$ result, the correlation deteriorates in $\Sigma F(E_t)$. The $E_t$ incorporates both wind and wave effects and represents the mechanical energy available for entraining air into water for the eventual SSA production. One would have expected $R_{FE}^2 > R_{FU}^2$, that is, an improvement in the correlation between $\Sigma F$ and $E_t$ over that between $\Sigma F$ and $U_{10}$, but instead the correlation dropped from $R_{FU}^2 = 0.72$ to $R_{FE}^2 = 0.57$ !

The counter-intuitive result is intriguing. With further examination, it turns out that the parameterization $\Sigma F(E_t)$ effectively sorts out the disparate data into several distinctive populations. In Fig. 16a we reproduce Fig. 13b in a slightly different fashion, separating the data into three main groups: (i) rising winds (red symbols), (ii) falling winds (blue symbols), and (iii) quasi-steady condition (green symbol), denoted as GR, GF and GQS, respectively. With reference to Fig. 16b, which is an abridged version of Fig. 11, the three groups are: GR: rising-wind events 2R and 4R; GF: falling-wind events 1F, 3F and 6F; and GQS: quasi-steady event 5QS; numbers 1 to 6 of the events increase chronologically. In Fig. 13b only measurements coinciding with $T_b$ are displayed, whereas in Fig. 16a all coexist SSA and $E_t$ are shown; the available observations more than doubled (increased from 128 to 288).

Several features become immediately apparent from the $\Sigma F(E_t)$ parameterization as presented in Fig. 16a: (a) GR, GF and GQS are well sorted out; and (b) There are clear time lags between $U_{10}$, $E_t$ and $\Sigma F$ and a general trend of increasing $\Sigma F$ for the same $E_t$ value as time advances. These are explained in further detail below.

*a. Wind steadiness*

The rates of $\Sigma F$ change as a function of $E_t$ are different in GR, GF and GQS. For this discussion, we consider the rate $\beta = d\Sigma F / dE_t$; conceptually $\beta$ probably can be regarded as a measure of the efficiency of wave breaking for SSA production. The magnitude of $\beta$ is high in the less energetic breaking condition ($E_t < 0.35$ W m$^{-2}$), to the left side of the dashed line in Fig. 16a. For $E_t > 0.35$ W m$^{-2}$, $\beta$ is noticeably lower in rising winds (GR, red symbols) than in falling winds (GF, blue symbols). Higher SSA flux in falling wind than that in rising winds is also reported by Ovadnevaite et al. (2012). In closely related whitecap coverage measurements, Sugihara et al. (2007) and Callaghan et al. (2008) also report higher whitecap fraction as a function of wind speed in falling winds than that in rising winds.

A plausible explanation of this observation is offered here. It indicates that all breakings are not created equal. With respect to SSA production, soft and gentle breaking as occurring in low or falling winds are much more efficient and can produce more SSA flux per unit of energy dissipation rate. In rising high winds, the more violent breaking events entrain larger air cavities that escape by buoyancy before they are fully broken down into smaller bubbles for more efficient SSA production. As a result, it costs more units of energy dissipation rate to produce the same quantity of SSA flux by the more-violent breaking in strong rising winds compared to the situation with gentler breakings in low or falling winds. So, efficiency cannot be rushed.

There are also alternative interpretations, of course. For example, we may connect rising and falling winds with developing and developed seas. In such terms, part of the $E_t$ for rising wind still goes to growing the waves, thus less of the $E_t$ goes into entraining air and producing SSA. For the falling wind, the wave field is developed so all the $E_t$ goes into pushing air into water. That is, less air is entrained during developing seas and more during developed seas. However, the fraction of momentum or energy needed to sustain wave growth is only a few percent of the wind input. For example, the analyses of Hasselmann et al. (1973), Donelan (1998) and Hwang and Sletten (2008) all give a low single-digit percentage in field conditions ($\omega_{\#} = U_{10}c_p^{-1}$ typically less than about 2). In comparison, the energy required to entrain the bubbles against buoyancy represents 30% to 50% of the total surface wave energy dissipated during breaking (Lamarre and Melville 1991). Hwang et al. (2012) present analysis results of wave growth in mixed seas and unsteady wind forcing. In terms of the dimensionless growth functions, the wave energy level for a given fetch or wave age is higher in ***both*** rising and falling phases of unsteady events compare to that in quasi-steady events. The difference in the wave growth is subtle although detectable. Given the great disparity of the energy levels required for sustaining wave growth and entraining air, it seems that the difference of wave development in rising and falling winds is not enough to explain the factor-of-two difference in the SSA production between falling and rising winds, as shown in Fig. 16a.

Support for the cavity size argument can be found in the research of length and velocity scales of breaking waves. Source function analysis of short scale waves shows that the length scale of the signature of dissipation function is shorter in mixed sea than that in wind seas as shown in Fig. 4 of Hwang and Wang (2004). Similarly, the breaking wave velocity scale is found to be smaller in mixed seas than that in wind seas based on feature tracking or Doppler analysis of sea spikes, which are closely related to breaking waves (Frasier et al. 1998; Hwang et al. 2008a, b). Because the buoyancy of the entrained air cavity is proportional to the volume, the smaller breaker length scale of mixed seas (more likely in falling winds) is more efficient for SSA or whitecap generation.

The rate $\beta$ in the quasi-steady event 5QS is rather erratic and signifies that in a natural setting, it is very unlikely to encounter truly steady wind conditions. In the situation of our experiment, the quasi-steady event is preceded by a couple of strong wind episodes and the wave system is composed of windsea superimposed on strong swell (Fig. 3a). The separation frequency between windsea and swell fluctuates wildly, resulting in a large variation of the estimated energy dissipation rate. The situation is further complicated by the time lags between $U_{10}$, $E_t$ and $\Sigma F$, which is discussed next.

*b. Event time lags*

The time series illustrated in Fig. 16b clearly show the response lags of $E_t$ and $\Sigma F$ with respect to the rising and falling of $U_{10}$. Generally, the separation between peaks of $U_{10}$ and $E_t$ are much closer than those of $U_{10}$ and $\Sigma F$. For example, in the quasi-steady segment (event 5QS), the peaks of $U_{10}$ and $E_t$ are within the same hour (0.72 h difference) whereas those of $U_{10}$ and $\Sigma F$ are 5.5 h apart. These response time lags represent one source of data scatter. Two examples are highlighted here.

(i) Using the quasi-steady event 5QS as an illustration, in Fig. 16a the red rectangle box contains the data points that $E_t$ is near the peak while $\Sigma F$ is still slowly evolving, and the blue rectangle box contains the data with $\Sigma F$ near its peak while $E_t$ is near the trough of this event.

(ii) Transitioning from rising event to falling event, the lags contribute to the appearance of increasing $\Sigma F$ under decreasing $E_t$, as shown in the magenta box in Fig. 16a containing the data points in the neighborhood between the $E_t$ peak and the $\Sigma F$ peak crossing from rising event 2R to falling event 3F. To aid the visualization, the temporal evolution of the data points in these two events are shown with connected symbols.

*c. The issue of temporal accumulation*

Summarizing the observations given in *(a)* and *(b)* above, the relationship $\Sigma F(E_t)$ as shown in Fig. 16a describes the evolution of episodic SSA production by breaking waves. In our experiment, the atmosphere starts out with very few SSA (minimum near time $t$=~117.3 d and again a local low at $t$=~118.1 d in Fig. 16b). When viewed in terms of $\Sigma F(E_t)$, the six identified events of wind fluctuations show a generally spiraling upward trend of $\Sigma F$ increasing its magnitude with time for a given level of energy dissipation rate $E_t$. This result appears to highlight the long time-constant of SSA: on the order of several hours for 10 μm particles and several days for 1 μm particles (e.g., Hoppel et al. 2002; Lewis and Schwartz 2004). It is tempting to interpret this result as that "over the period of about 4 days, the SSA still has not reached steady state and the data exhibits significant temporal variations." This long-time-constant interpretation is premature, however, because of the large fluctuations of the RH, especially in events 1F and 2R compared to the later periods (see Fig. 11e). In such a situation, it is important to make the distinction between the total atmospheric aerosol loading within a certain "*in situ*" size band (the $\Sigma F$, $\Sigma N$, and $\Sigma V$ presented up to this point) and the aerosol properties "*at the source*" of a certain size band. Indeed, the relative humidity is a key meteorological parameter in aerosol models (e.g., Fitzgerald 1978; Gathman 1983; Gerber 1985).

*d. Humidity effect*

Figure 17 shows four SSA size spectra, each the mean over 2 consecutive hours of the 20-min average measurements. These four spectra are selected, two each, in rising- and falling-wind periods with varying RH; the average wind speeds are all within $11.0\pm0.1$ m s$^{-1}$. It is clear that at a lower ambient RH, the measured $r_{is}$ corresponds to a larger source radius $r_{98} \approx 2r_{80}$ (e.g., Lewis and Schwartz 2004). Because the SSA number and flux in the coarse mode decreases exponentially with size (Fig. 17a; also Figs. 7b and 7d), the smaller size particles missed in the lower RH data contributes significantly to the lower integrated SSA values in the earlier part of the experiment, as shown in the data during $t$=118.1~119.23 d (Fig. 13) or event 2R (Fig. 16a).

We can compare the black and magenta curves of the two rising wind events 2R and 4R in Fig. 17, with average RH=71.4% and 89.3%, respectively. Although the black curve (2R, lower RH) is generally higher than the magenta curve, because of its reduced coverage of the small size in the generation source, the integrated SSA number and flux within the same $r_{is}$ range are in fact much smaller compared to those of the magenta curve (4R, higher RH): [$\Sigma N$ (m$^{-3}$), $\Sigma F$ (m$^{-2}$ s$^{-1}$)] = [1.01×10$^6$, 1.98×10$^4$] for 2R and [2.43×10$^6$, 3.44×10$^4$] for 4R. Interestingly, the integrated volumes $\Sigma V$ (μm$^3$ m$^{-3}$) are much closer: 1.41×10$^8$ vs. 1.24×10$^8$. The result suggests that the SSA mass or volume at the generation source is similar in the two events although they are 1.3 days apart, and the initial impression of temporal accumulation as might have been deduced from Fig. 16a or 13 is more likely caused by the much lower RH in the earlier event, thus the *in situ* size range missed the smaller particles at the generation source.

On the other hand, the two falling wind events (green and blue curves in Fig. 17), occur 1.8 days apart with average RH=81.7% and 84.6%, have similar size coverage of the source SSA particles, but the SSA properties in terms of number, volume and flux all increase with time for these two examples. To have a more comprehensive comparison, Fig. 18 presents the complete data set, showing the flux and volume integrated over the size range $r_{80} = 2$ to 10 μm to avoid the mismatch of the SSA size range at the generation source caused by the RH fluctuation. On the top row they are plotted against $E_t$ and on the bottom row they are against $U_{10}$. Although data scatter is large, the impression of temporal accumulation derived from Fig. 16a disappears; the most prominent feature left behind is the contrast between rising winds and falling winds. Indeed, the parameterization with $E_t$, which considers the wind and wave evolution, is useful for elucidating the physics of SSA production. The magnitudes of $\Sigma F$ and $\Sigma V$ in falling winds are about twice those in rising winds for the same $E_t$, indicating very different wave breaking efficiency for SSA production as discussed in Sec. 4.1.a.

As an additional comment on the time evolution issue, using $t$=118.1 d, when the SSA value is very small, as the reference starting time, the results illustrated in Fig. 18 indicate that data collected several hours after the starting time is not very different from that collected days later (compare events 2R and 4R, or 3F and 6F). This is in contrast to the notion that the time constant of particles with radii of several μm is on the order of days based on the consideration of the particle settling velocity alone (e.g., Hoppel et al. 2002; Lewis and Schwartz 2004).

To reiterate, in addition to the physical reasons given in Sec. 4.1.a-b, the requirements of the dry deposition method for estimating the SSA production can potentially be suggested as the cause of the differences between rising and falling winds. As one of its assumptions, dry deposition method requires steady state conditions that persisted sufficiently long time to saturate marine boundary layer with aerosol particles and thus establish a balance between upward and downward fluxes. However, any change in wind speed and thus effective aerosol production rate disrupts this balance, leading to underestimates of surface flux in rising winds and overestimates of surface flux in falling winds. This property can be suggested as the reason for the delay in production flux observed in Fig 16a. However, if this were the case, the surface flux's dependence on the brightness temperature shown in Fig. 13d would demonstrate similar delayed behavior as its dependence on the dissipation rate (Figs. 13b, 16a), which it does not. Therefore, the ability of the microwave radiometer to capture the difference in SSA production rate in falling and rising winds eliminates the shortcomings of the dry deposition method as the primary cause and supports a physical reasoning for the observed phenomenon, such as the breaking efficiency given in section 4.1.a. This leads to our last discussion item: the ocean surface roughness.

### 4.2. Remote sensing of ocean surface roughness

Figure 13 shows a much improved correlation in $\Sigma F(\Delta T_b)$ compared to $\Sigma F(U_{10})$, $\Sigma F(E_t)$ or $\Sigma F(P_{<f>})$. In Fig. 19, the SSA flux integrated over the size range $r_{80} = 2$ to 10 µm is presented in the same format as Fig. 13 and the conclusion is not altered regarding the improved correlation between $\Sigma F_{2\sim10}$ and the roughness component of brightness temperature $\Delta T_b$ compared to the other three forcing factors that are closely connected to wave breaking. Whereas the $E_t$, and to a different degree $U_{10}$ or $P_{<f>}$, parameterization sorts the SSA data into distinct falling wind and rising wind groups (Figs. 18, 19), the $\Delta T_b$ parameterization blends them together more smoothly. In essence: "Go short" when searching for single-variable parameterizations.

In the discussion of Fig. 13, we have ventured an explanation that although wave breaking is an important generation source, the final state of the SSA in the marine boundary layer is controlled by the turbulent processes that are closely connected to the ocean surface roughness. The time scale of turbulent mixing is much shorter than the time scale of SSA particle settling velocity, and provides a favorable condition for the application of dry deposition method to calculate the SSA flux. Although the exact processes are still to be identified, it is noteworthy that similar results showing much better correlation with roughness ( MSS) than wind speed have been reported for the gas transfer velocity (e.g., Bock et al. 1999; Frew et al. 2004, 2007), as has been described in the Introduction section.

Microwave sensors are especially suited for obtaining the ocean surface roughness properties. In S14 a parameterization model using the passive microwave radiometer output is presented. Active microwave sensors such as altimeter, scatterometer and synthetic aperture radar (SAR) also yield ocean surface roughness information. Frew et al. (2007) describe an application of dual-frequency (Ku and C band) spaceborne altimeter data for global estimation of air-sea gas transfer velocity fields. The application can be extended to other airborne or spaceborne scatterometer or SAR data. Also, ocean surface backscattering from radars installed on land, towers or ships contains critical information of the ocean surface roughness. In particular, the shipborne X band navigational radar backscattering onboard field cruises can be useful supplemental data even though it is not calibrated: post processing comparing the uncalibrated average radar cross section (RCS) as a function of $U_{10}$ with the established calibrated NRCS (normalized RCS) curves would render the supplemental data quite valuable for retrieving the small scale ocean surface roughness.

## 5. Summary

This paper presents the analysis results of simultaneous measurements of aerosol, wind, wave, underwater acoustic noise and microwave brightness temperature collected during a 2012 experiment conducted onboard the free-drifting FLIP. Focusing on the wave breaking effects on the SSA production, the analysis casts parameterizations of total (size integrated) SSA properties -- namely, number ($\Sigma N$), volume ($\Sigma V$), and flux ($\Sigma F$) -- in terms of the wind speed $U_{10}$, breaking wave energy dissipation rate $E_t$ combining the wind and wave variables, and acoustic noise $P_{<f>}$ attributable to bubbles generated by breaking waves. The brightness temperature data is processed to extract the roughness and foam components, of which the roughness component $\Delta T_b$ is found to be more correlated to the SSA properties. The forcing factors $E_t$, $P_{<f>}$, and $\Delta T_b$ represent ocean surface processes with length scales of O(10 m), O(1 m), and O(0.01 m), respectively.

Using the $U_{10}$ parameterization as a reference, the correlation coefficients of the SSA properties with $E_t$ deteriorate. This is unexpected considering that $E_t$ incorporates both wind and wave properties. It represents the available mechanical energy discharged by wave breaking for entraining air into water and for the subsequent SSA production. Further analysis indicates that although $E_t$ is not a good candidate for single-variable parameterization, it is very useful for clarifying the efficiency of wave breaking regarding SSA production. In particular, except in low-dissipation breaking conditions, the SSA production in rising wind events is less effective than that in falling winds. Our interpretation of this result is that in strong rising winds, the breaking is likely more violent and entrains larger air cavities that do not stay in water long enough to be effectively broken down to smaller bubbles for more efficient SSA generation. In contrast, the gentler breaking in low or falling winds is able to make better use of the smaller entrained air cavities because they stay underwater longer for turbulence splitting. Support of the air cavity size argument can be found in the analyses of breaking wave length and velocity scales showing smaller breaker size in mixed seas than in wind seas (Frasier et al. 1998; Hwang et al. 2008a, b), and the source function analysis of short waves indicating that the length scale of the signature of dissipation function shifts to shorter scale in mixed seas compared to that in wind seas (Hwang and Wang 2004).

The correlation analysis of the SSA properties ($\Sigma N$, $\Sigma V$, and $\Sigma F$) further shows a persistent improvement toward the shortest-scale parameterization. This is interpreted as that whereas wave breaking represents an important source for SSA generation, the final state of atmospheric aerosol properties is determined by turbulent transport processes that are better characterized by the ocean surface roughness. Active and passive microwave sensors are especially suitable for obtaining the ocean surface roughness properties. They represent a valuable resource to aid our investigation of the complicated air-sea interaction problems.

**Acknowledgements**

This work is sponsored by the Office of Naval Research (NRL program element 61153N, work units 4641, 4500 and 4278). We appreciate the funding support for our use of FLIP by Robert Schnoor (Naval Research Facilities Program at ONR), and by Joan Gardner and Edward Franchi (NRL Platform Support Program). Captain William Gaines, FLIP program manager at Marine Physics Laboratory (MPL) at Scripps Institution of Oceanography, was indispensable in organizing the field campaign. We appreciate Tom Golfinos, Officer-in-Charge for FLIP, and crew members Johnny, Dave, Frank, and Jerry for their hard work, endurance, and camaraderie. We would also like to thank George Trekas and his colleagues at the MPL Machine Shop for their expertise and skills in devising the instrument deployment on the FLIP booms.

**Appendix: Supplemental Information**

**A. *Foam and roughness components of brightness temperature***

There are two main components of the wind and wave contribution to the deviation of the brightness temperature from the flat surface value: foam and roughness. The formula connecting $T_b$ and $T_s$ can be written as (Hwang 2012)

$$T_{bp} = \left( e_{0p} + \delta e_{fp} + \delta e_{rp} \right) T_s \,, \tag{A1}$$

where subscript $p$ is polarization, $e_0$ is the emissivity of the flat surface without foam, $\delta e_f$ is the emissivity correction due to foam for the flat surface contribution, and $\delta e_r$ is the rough surface contribution. These three components are given by:

$$e_{0p} = 1 - \left| R_{pp}^{(0)} \right|^2 \,, \tag{A2}$$

where $R_{pp}^{(0)} \left( \varepsilon, \theta \right)$ is the Fresnel reflection coefficient of polarization $p$, the value is determined by the incidence angle $\theta$ and the frequency-dependent relative permittivity $\varepsilon$;

$$\delta e_{fp} = e_{ep} - e_{0p} \,, \tag{A3}$$

where $e_{ep}$ is the effective emissivity of the flat surface with foam (to be further discussed in the next paragraph); and

$$\delta e_{rp} = \int_0^\infty \int_{-\pi}^\pi W \left( k', \phi' \right) g_p \left( f, \theta, \phi, \varepsilon_e, k', \phi' \right) k' \, d\phi' \, dk' \,, \tag{A4}$$

where $W$ is the directional ocean surface roughness spectrum with $k$ the surface wave number and $\phi$ the propagation direction, and $g_p$ is the electromagnetic "weighting" function for the $p$ polarization; the full expression of $g_p$ is given by Yueh et al. (1994a, Appendices 1-3) and Johnson and Zhang (1999, p. 2308). An effective relative permittivity $\varepsilon_e$ is used by Hwang (2012) for evaluating (A4), instead of the sea water relative permittivity $\varepsilon_{sw}$ (without foam adjustment) in the original papers by Johnson and Zhang (1999) and Yueh et al. (1994a).

The relative permittivity of air and water differ considerably for microwave frequencies, therefore even a small amount of air can produce non-trivial changes in the relative permittivity of the resulting mixture. To account for the effect of entrained air, ideally we need to know the fraction of air in water (void fraction) and the vertical distribution of the bubble clouds carrying the air into water. Considering the small penetration depth of microwaves, for example, the skin depth is about 2 mm at 10 GHz (Plant 1990), Hwang (2012) uses the fraction of whitecap coverage as a proxy of the void fraction in the mixing rule for the evaluation of the effective relative permittivity $\varepsilon_e$. The approach renders the emission problem of foam-covered ocean surface to quasi-2D (horizontal). The whitecap fraction represents an upper bound of the void fraction because it is equivalent to assuming 100% of air in the depth of microwave influence under the foamy area, whereas the air entrainment decreases exponentially with water depth (e.g., Wu 1981; Hwang et al. 1990). The effective relative permittivity $\varepsilon_e$ of the air-water mixture is computed with the quadratic mixing rule in a similar approach of Anguelova (2008):

$$\varepsilon_e = \left[ f_a \varepsilon_a^{1/2} + \left(1 - f_a\right) \varepsilon_{sw}^{1/2} \right]^2, \tag{A5}$$

where $\varepsilon_{sw}$ is relative permittivity of sea water without whitecaps and $\varepsilon_a = 1$ is relative permittivity of air, and $f_a$ is air fraction approximated by the whitecap fraction.

Based on this "equivalent medium" approach, the calculated results are in very good agreement with a global dataset of WindSat microwave radiometer measurements with wind speed coverage up to about 42 m s$^{-1}$ (Meissner and Wentz 2009). The WindSat measurements include five microwave frequencies (6, 10, 18, 23 and 37 GHz) for both vertical and horizontal polarizations; see Fig. 3 in Hwang (2012) and Fig. 9 in Hwang et al. (2013).

## B. Surface wave modulation of underwater acoustic noise

The phase center of our acoustic array is at 18.2 m underwater and the significant wave height reaches to more than 4 m during the high wind period. It is thus of some concern about the quantitative value of the underwater acoustic noise normalized by the area of reception, because the surface area intercepted by the cone of acoustic reception is obviously modulated by the surface waves passing over the acoustic array. Because there is no pressure sensor in the acoustic array and the surface wave sensor system and the acoustic array are not collocated, we are not able to do a point-by-point correction of the surface wave modulation. Only the statistical average is investigated here.

Let's define $A$ as the circular surface area of acoustic reception. In the absence of waves, it is related to the depth of acoustic sensor $h$ by

$$A = \pi R^2 = \pi \left( h \tan \theta \right)^2, \tag{A6}$$

where $R$ is the radius of the circle and $\theta$ is the angle between the vertical axis of the array and a point on the reception circle. Installed on the stable FLIP, the instantaneous depth of the array in the presence of wave motion is $h + \eta$, thus the instantaneous acoustic reception area is modulated as

$$A_\eta = \pi \left[ (h + \eta) \tan \theta \right]^2, \tag{A7}$$

The average ratio of the modulated area to the reference area in static condition can be written as

$$\left\langle \frac{A_\eta}{A} \right\rangle = \int_{-\infty}^{\infty} \left( 1 + \frac{\eta^2}{h^2} \right) p(\eta) d\eta, \tag{A8}$$

where $p(\eta)$ is the probability density function of surface elevation. The integral in (A8) can be rewritten as

$$\int_{-\infty}^{\infty}\left(1+\frac{\eta^2}{h^2}\right)p(\eta)\,d\eta = 1+\frac{1}{h^2}\int_{-\infty}^{\infty}\eta^2 p(\eta)\,d\eta \,. \tag{A9}$$

The integral on the right hand side of (A9) is, of course, the definition of the variance of the surface wave elevation, therefore,

$$\left\langle \frac{A_\eta}{A} \right\rangle = 1+\frac{\eta_{rms}^2}{h^2}\,. \tag{A10}$$

The maximum significant wave height during our experiment is 4.45 m, so the maximum $\eta_{rms}=H_s/4=1.11$ m; the sensor depth is 18.2 m and $\left\langle A_\eta A^{-1} \right\rangle$ never exceeds 1.004.

### C. Humidity adjustment of SSA size

At the source region near the air-sea interface, the RH for the SSA particles is about 98% with sea water of 35 psu salinity (Lewis and Schwartz 2004). Ejected into air with lower RH the particle size decreases through evaporation. As the atmospheric RH fluctuates, the particle size also fluctuates through absorbing moisture in higher RH or evaporation in lower RH environment. The thermodynamic processes of deliquesce and effloresces of aerosol particles are complicated and strongly dependent on the chemical compositions of the aerosols (e.g., Winkler 1973; Fitzgerald 1978; Tang et al. 1997; Lewis and Schwartz 2004). For example, the growth ratio $r_{is}r_{dry}^{-1}(RH)$ of maritime aerosol is much larger than that of continental aerosol.

Many useful approximation formulas for sea salt aerosols have been developed to convert the *in situ* measured size $r_{is}$ with the ambient RH input to the reference size $r_{80}$ at RH=80% (e.g., Lewis and Schwartz 2004). We have used the Gerber (1985) formulation (in a lookup table) for getting the dry particle size $r_{dry}$ in our data processing, but in order to make comparison with published results using $r_{80}$, we employ the approximation formulas $r_{80}=2r_{dry}$ and $r_{98}=4r_{dry}$ (Lewis and Schwartz 2004).

Figure A1a plots the ratios $r_{dry}(r_{80})$ and $r_{dry}(r_{98})$ for the coarse mode SSA reported here, showing reasonably good agreement between the lookup table (Gerber 1985) and the approximation formulas (Lewis and Schwartz 2004).

The lookup table at the early stage of our data processing lists triplets of [ $r_{is}$, RH, $r_{dry}$ ], it is difficult to obtain $r_{80}$ from the table directly; the additional convoluted interpolations may further introduce errors. Our approach of blending the two different schemes of RH correction for the SSA radius, i.e., using the Gerber (1985) algorithm in lookup table to obtain $r_{dry}$ from $r_{is}$, and Lewis and Schwartz (2004) approximation to obtain $r_{80}$ from $r_{dry}$, is not expected to create large errors with respect to the conclusions discussed in this paper regarding the physical processes impacting the SSA production. The new lookup table lists quadruplets [ $r_{is}$, RH, $r_{dry}$, $r_{80}$ ]. The ratio $r_{dry}r_{80}^{-1}$ based on Gerber (1985) ranges between 0.455 and 0.493 for RH between 70% and 98% (Fig. A1b).

## List of Tables

Table 1. Power-law coefficient of SSA size spectra: $X(r, U_{10}) = A_X(r)U_{10}^{a_x(r)}$, where $X$ can be $N$,

$V$ or $F$.

| $r_{80}$ (μm) | 2.00E+00 | 4.00E+00 | 6.00E+00 | 8.00E+00 | 1.00E+01 | 1.20E+01 | 1.40E+01 |
|---|---|---|---|---|---|---|---|
| $A_N$ | 5.52E+04 | 2.21E+04 | 3.46E+03 | 1.60E+02 | 4.50E+00 | 7.45E-01 | 1.50E-01 |
| $a_N$ | 1.21E+00 | 1.23E+00 | 1.69E+00 | 2.44E+00 | 3.45E+00 | 3.81E+00 | 3.85E+00 |
| $A_V$ | 2.53E+06 | 6.30E+06 | 3.39E+06 | 3.11E+05 | 2.03E+04 | 1.46E+03 | 1.48E+02 |
| $a_V$ | 1.07E+00 | 1.18E+00 | 1.62E+00 | 2.42E+00 | 3.37E+00 | 4.27E+00 | 4.74E+00 |
| $A_{F_{is}}$ | 2.42E+01 | 3.46E+01 | 1.42E+01 | 1.02E+00 | 5.06E-02 | 1.15E-02 | 3.44E-03 |
| $a_{F_{is}}$ | 1.31E+00 | 1.40E+00 | 1.82E+00 | 2.62E+00 | 3.59E+00 | 3.97E+00 | 3.98E+00 |
| $A_{F_U}$ | 1.35E+02 | 6.84E+01 | 1.48E+01 | 9.49E-01 | 4.13E-02 | 5.37E-03 | 8.40E-04 |
| $a_{F_U}$ | 2.17E+00 | 2.12E+00 | 2.49E+00 | 3.15E+00 | 4.03E+00 | 4.54E+00 | 4.72E+00 |

Table 2. Polynomial fitting coefficients of the SSA size spectra power-law coefficients: $Y(Z) = p_{Z1}Z^3 + p_{Z2}Z^2 + p_{Z3}Z + p_{Z4}$, where Z can be $A_N$, $a_N$, $A_V$, $a_V$, $A_F$ or $a_F$.

| $Z$ | $p_{Z1}$ | $p_{Z2}$ | $p_{Z3}$ | $p_{Z4}$ |
|---|---|---|---|---|
| $A_N$ | 1.03E+00 | -9.68E+00 | 1.60E+01 | 4.06E+00 |
| $a_N$ | -5.46E-01 | 3.10E+00 | -4.65E+00 | 2.11E+00 |
| $A_V$ | -1.36E+00 | 9.86E-01 | 4.24E+00 | 1.18E+01 |
| $a_V$ | -3.28E-01 | 2.11E+00 | -3.19E+00 | 1.38E+00 |
| $A_{F_{is}}$ | 9.87E-01 | -9.40E+00 | 1.75E+01 | -4.81E+00 |
| $a_{F_{is}}$ | -4.73E-01 | 2.70E+00 | -4.00E+00 | 1.91E+00 |
| $A_{F_U}$ | 4.18E-02 | -4.81E+00 | 9.40E+00 | 6.43E-01 |
| $a_{F_U}$ | -4.73E-01 | 2.70E+00 | -4.00E+00 | 1.91E+00 |

**List of Figures**

Fig. 1. The FLIP track during the BREWEX in Pacific Ocean about 200 km west of California coast. The dark segment marks the period of our main interest: Day 117 to 122. The locations of 4 NDBC buoys are also shown.

Fig. 2. Comparisons of wind and wave measurements onboard FLIP with those of four buoys identified in Fig. 1: (a) Wind speed $U_{10}$; (b) Wind direction $\theta_U$; (c) Significant wave height $H_s$; and (d) Spectral peak wave period $T_p$.

Fig. 3. Sea state, swell and windsea separation, and windsea energy dissipation computation: (a) Temporal variation of the wave spectrum, for reference, $U_{10}$ and $f_{pw}$ are superimposed; (b) The significant wave heights of the windsea and swell components: $H_{sw}$ and $H_{ss}$, respectively; and (c) The spectral peak periods of the windsea and swell components: $T_{pw}$ and $T_{ps}$, respectively; (d) Significant wave height $H_s$ and windsea energy dissipation $E_t$; and (e) Windsea peak phase speed $c_{pw}$, wind speed $U_{10}$ and inverse wave age $\omega_{\#} = U_{10}c_p^{-1}$.

Fig. 4. Processed results of the 10.7 GHz microwave brightness temperature $T_b$ measurements: (a) Time series of wind speed $U_{10}$, and the $T_b$ deviation $\delta T_{bp}$ (from the flat surface value), subscript $p$ is polarization (H: horizontal, V: vertical). (b) The $\delta T_{bp}$ dependence on wind speed. The modeled curves (Hwang 2012) separating the foam and roughness contributions and the sum are illustrated for comparison. (c) The difference $\Delta T_b = \delta T_{bH} - \delta T_{bV}$ is dominated by the roughness contribution; the model curves are illustrated for comparison.

Fig. 5. The underwater acoustic noise in the frequency range between 1250 and 2350 Hz: (a) Frequency spectra in different wind speeds; (b) The proportionality coefficient $A_f\left(U_{10}\right)$; and (c) Exponent $a_f\left(U_{10}\right)$, of the frequency spectrum expressed as $P\left(f;U_{10}\right) = A_f f^{a_f}$.

Fig. 6. The underwater acoustic noise in the frequency range between 1250 and 2350 Hz: (a)

Spectral component dependence on wind speed; (b) The proportionality coefficient $A_U(f)$; and (c) Exponent $a_U(f)$, of the frequency spectrum expressed as $P(U_{10}; f) = A_U U_{10}^{a_U}$.

Fig. 7. The SSA size spectra: (a) Volume d$V$/dln$r$; (b) Number d$N$/dln$r$; (c) Flux computed with (3) and (4): d$F_{is}$/dln$r$; and (d) Flux computed with (5) and (6): d$F_U$/dln$r$.

Fig. 8. Power-law relationship of the SSA size spectral components: (a) Volume d$V$/dln$r$; (b) Number d$N$/dln$r$; (c) Flux computed with (3) and (4): d$F_{is}$/dln$r$; and (d) Flux computed with (5) and (6): d$F_U$/dln$r$.

Fig. 9: The coefficients of power-law empirical model functions for the SSA size spectra, $dN/d\ln r$, $dV/d\ln r$, $dF_{is}/d\ln r$, and $dF_U/d\ln r$: (a) proportionality coefficient $A$, and (b) exponent $a$.

Fig. 10: Empirical model function (EMF) of d$F_U$/dln$r$, and its comparison with data and other model functions (Monahan et al. 1986: M86; Smith et al. 1993: S93; Lewis and Schwartz 2004: L04). Plotting symbols circle, plus and triangle show data at 5, 10 and 15 m s$^{-1}$, respectively; solid, dashed, and dashed-dotted curves show various models (described in the legend with different colors) at 5, 10 and 15 m s$^{-1}$, respectively.

Fig. 11. Combined time series of underwater acoustic noise $P$, SSA flux $\Sigma F$, windsea energy dissipation rate $E_t$, roughness component of brightness temperature deviation $\Delta T_b$, and wind speed $U_{10}$. The time series are shown in linear scale in separate panels (a)-(e) on the left column; and together in logarithmic scale on the right column (f). Both $P_{2350}$ and $P_{<f>}$ are shown in (a); and both $\Sigma F_{is}$ and $\Sigma F_U$ are given in (b).

Fig. 12. Scatter plots showing the power-law dependence on $U_{10}$ for $P$, $\Sigma F$, $E_t$, and $\Delta T_b$: (a) Data unsorted; and (b) Sorting of 4 time segments illustrated in different colors. See text for more detail.

Fig. 13. Correlation between SSA flux $\Sigma F$ and (a) Wind speed $U_{10}$; (b) Windsea energy dissipation rate $E_t$; (c) Underwater acoustic noise $P_{<f>}$; and (d) Roughness component of brightness temperature deviation $\Delta T_b$; only co-registered data are shown (constrained mainly by $\Delta T_b$).

Fig. 14. Comparison of different SSA representations and their dependence on $U_{10}$ and $\Delta T_b$: (a) $\Sigma N(U_{10})$; (b) $\Sigma V(U_{10})$; (c) $\Sigma N(\Delta T_b)$; and (d) $\Sigma V(\Delta T_b)$; only co-registered data are shown (constrained mainly by $\Delta T_b$).

Fig. 15. Same as Fig. 14 but for dependence on $E_t$ and $P_{<f>}$.

Fig. 16. (a) Episodic behavior of the SSA flux showing differences in the rate of change as a function of $E_t$ in rising wind, falling wind and quasi-steady conditions; and (b) Time series of $U_{10}$, $E_t$ and $\Sigma F$, with 6 episodes identified. The general trend of rising and falling winds in $\Sigma F(E_t)$ is indicated by the arrows in (a); effects of time lags between $U_{10}$, $E_t$ and $\Sigma F$ are illustrated with dotted lines connecting 2R and 3F events and the magenta box highlighting the transition segments, as well as the red and blue boxes of the 5QS event; see text for more discussion.

Fig. 17. Relative humidity effect on the SSA size spectra; here 2-h average spectra in 2 rising wind events and 2 falling wind events are illustrated, the average wind speeds are within $11.0\pm0.1$ m s$^{-1}$: (a) d$F$/dln$r$, and (b) d$V$/dln$r$.

Fig. 18. Episodic behavior of the SSA production showing differences in the rate of change (in rising wind, falling wind and quasi-steady conditions) in terms of: (a) $\Sigma F_{2\sim10}(E_t)$; (b) $\Sigma V_{2\sim10}(E_t)$; (c) $\Sigma F_{2\sim10}(U_{10})$; and (d) (a) $\Sigma V_{2\sim10}(U_{10})$, where subscript $2\sim10$ denotes integration over the size range $r_{80}$=2 to 10 μm.

Fig. 19. Same as Fig. 13 but for integration over the size range $r_{80}$=2 to 10 μm.

Fig. A1. (a) Comparison of two RH correction schemes of SSA radius used in this paper: curves

(Gerber 1985) and symbols (Lewis and Schwartz 2004); (b) The ratio $r_{dry}r_{80}^{-1}$ as a function of

RH based on Gerber (1985).

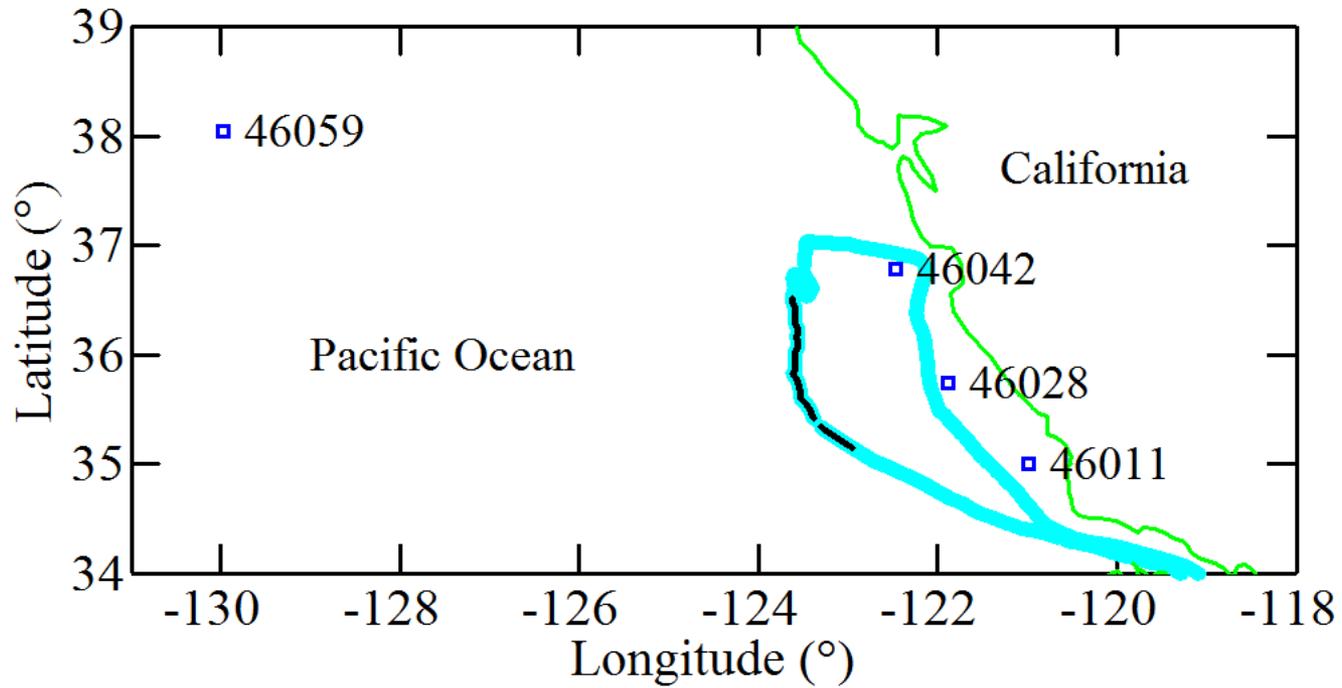

Fig. 1. The FLIP track during the BREWX in Pacific Ocean about 200 km west of California coast. The dark segment marks the period of our main interest: Day 117 to 122. The locations of 4 NDBC buoys are also shown.

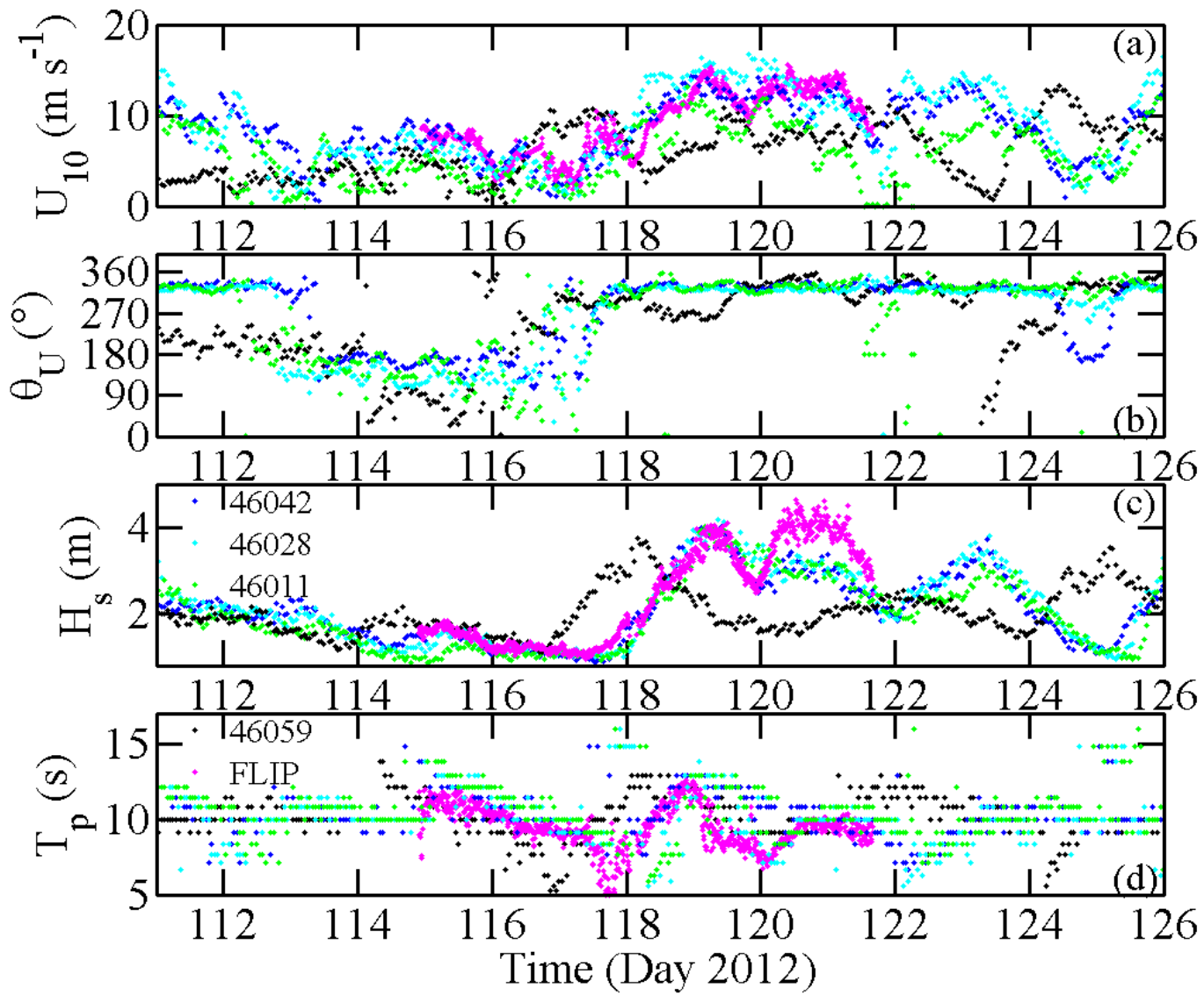

Fig. 2. Comparisons of wind and wave measurements onboard FLIP with those of four buoys identified in Fig. 1: (a) Wind speed $U_{10}$; (b) Wind direction $\theta_U$; (c) Significant wave height $H_s$; and (d) Spectral peak wave period $T_p$.

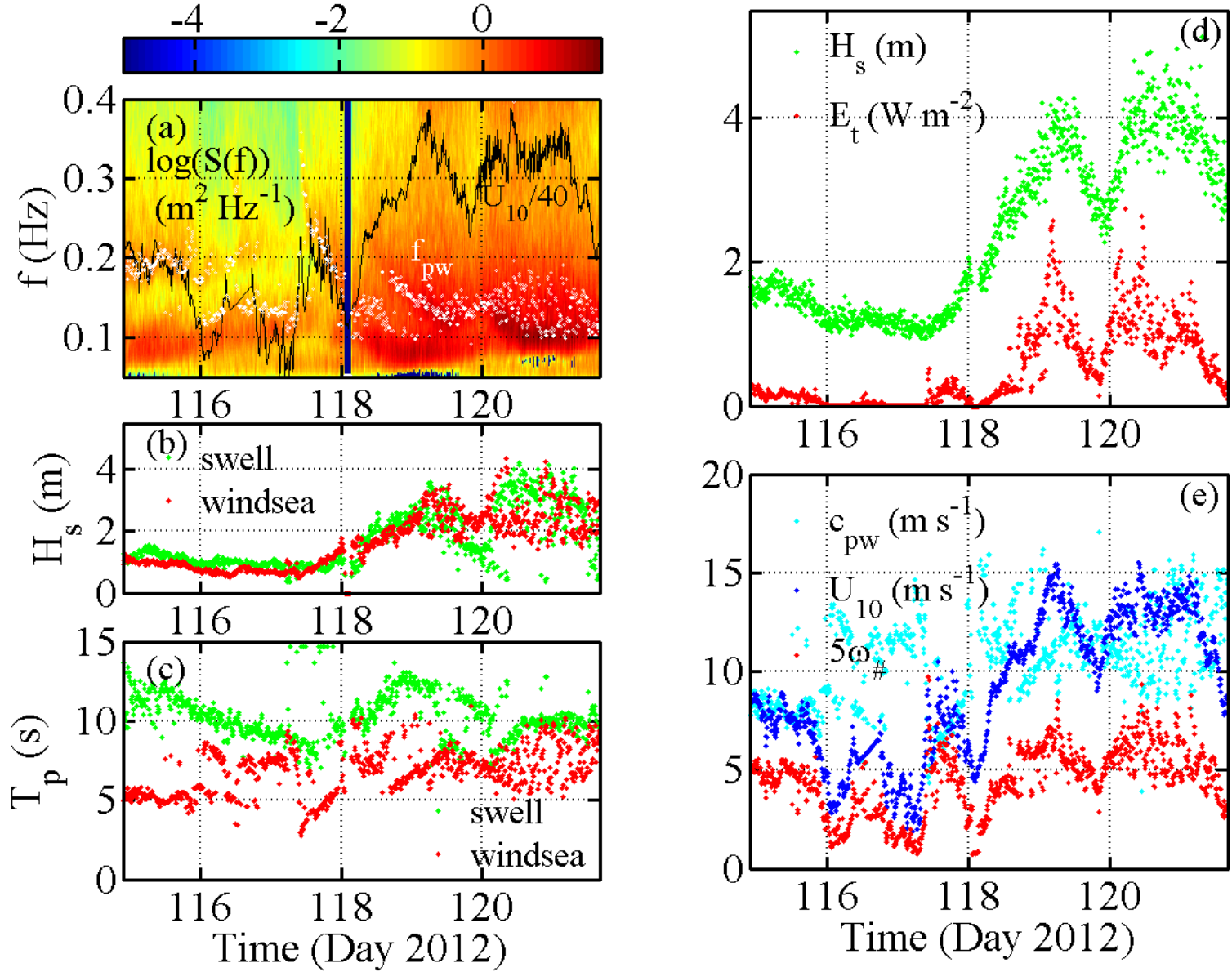

Fig. 3. Sea state, swell and wind-sea separation, and wind-sea energy dissipation computation: (a) Temporal variation of the wave spectrum, for reference, $U_{10}$ and $f_{pw}$ are superimposed; (b) The significant wave heights of the wind-sea and swell components: $H_{sw}$ and $H_{ss}$, respectively; and (c) The spectral peak periods of the Wind-sea and swell components: $T_{pw}$ and $T_{ps}$, respectively; (d) Significant wave height $H_s$ and wind-sea energy dissipation $E_t$; and (e) Wind-sea peak phase speed $c_{pw}$, wind speed $U_{10}$ and inverse wave age $\omega_\# = U_{10}c_p^{-1}$.

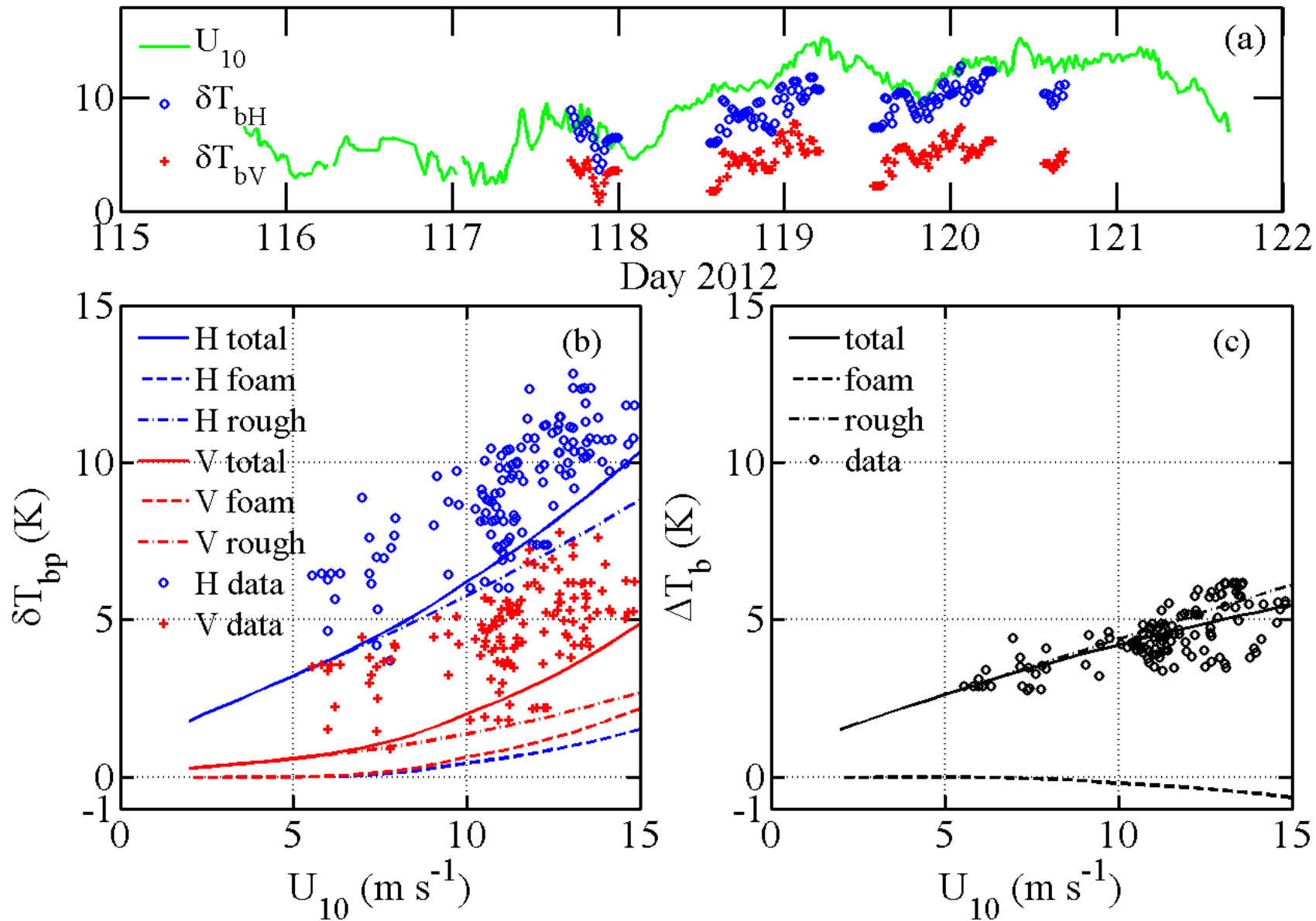

Fig. 4. Processed results of the 10.7 GHz microwave brightness temperature $T_b$ measurements: (a) Time series of wind speed $U_{10}$, and the $T_b$ deviation $\delta T_{bp}$ (from the flat surface value), subscript $p$ is polarization (H: horizontal, V: vertical). (b) The $\delta T_{bp}$ dependence on wind speed. The modeled curves (Hwang 2012) separating the foam and roughness contributions and the sum are illustrated for comparison. (c) The difference $\Delta T_b = \delta T_{bH} - \delta T_{bV}$ is dominated by the roughness contribution; the model curves are illustrated for comparison.

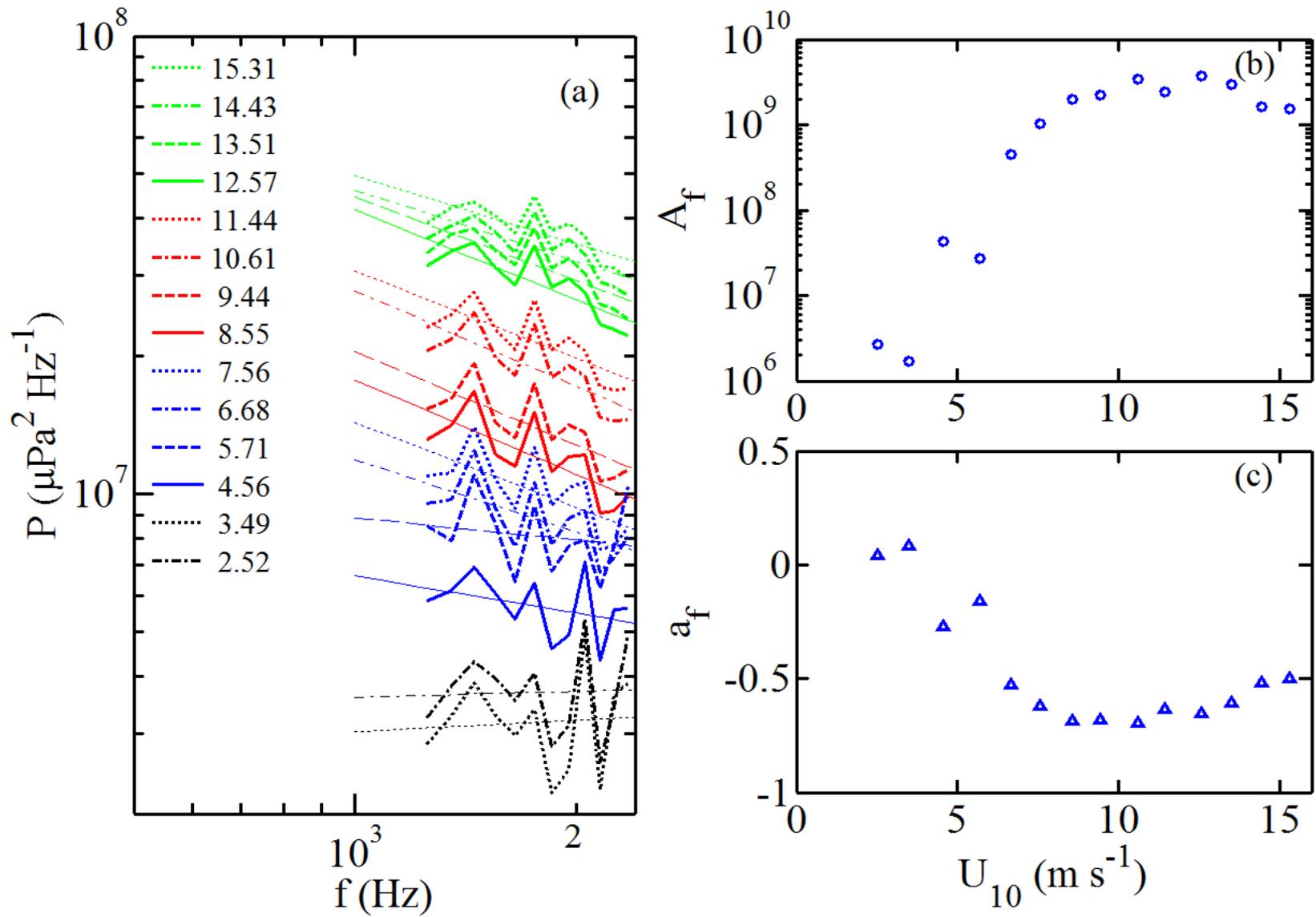

Fig. 5. The underwater acoustic noise in the frequency range between 1250 and 2350 Hz: (a) Frequency spectra in different wind speeds; (b) The proportionality coefficient $A_f(U_{10})$; and (c) Exponent $a_f(U_{10})$, of the frequency spectrum expressed as $P(f;U_{10}) = A_f f^{a_f}$.

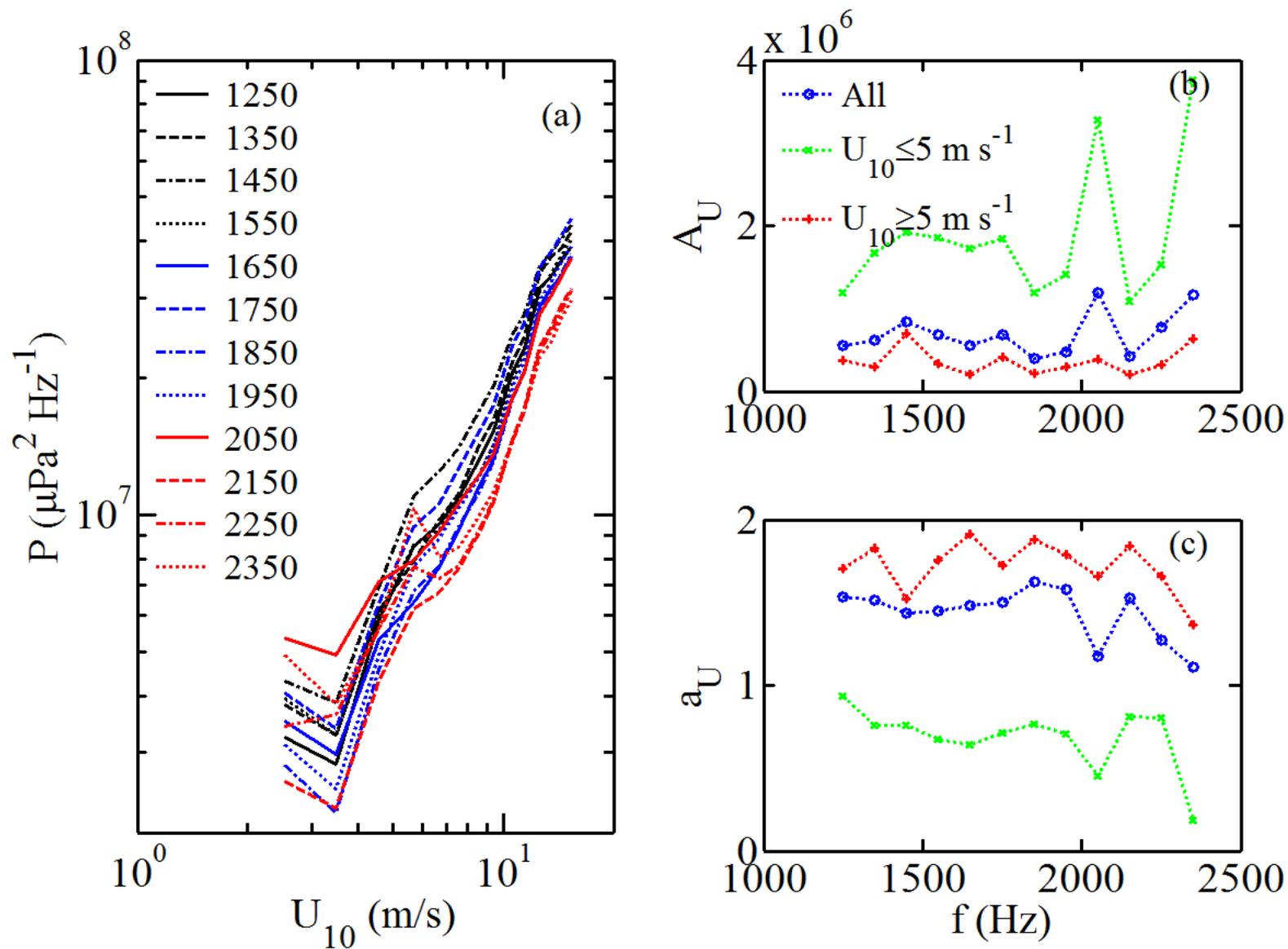

Fig. 6. The underwater acoustic noise in the frequency range between 1250 and 2350 Hz: (a) Spectral component dependence on wind speed; (b) The proportionality coefficient $A_U(f)$; and (c) Exponent $a_U(f)$, of the frequency spectrum expressed as $P(U_{10}; f) = A_U U_{10}^{a_U}$.

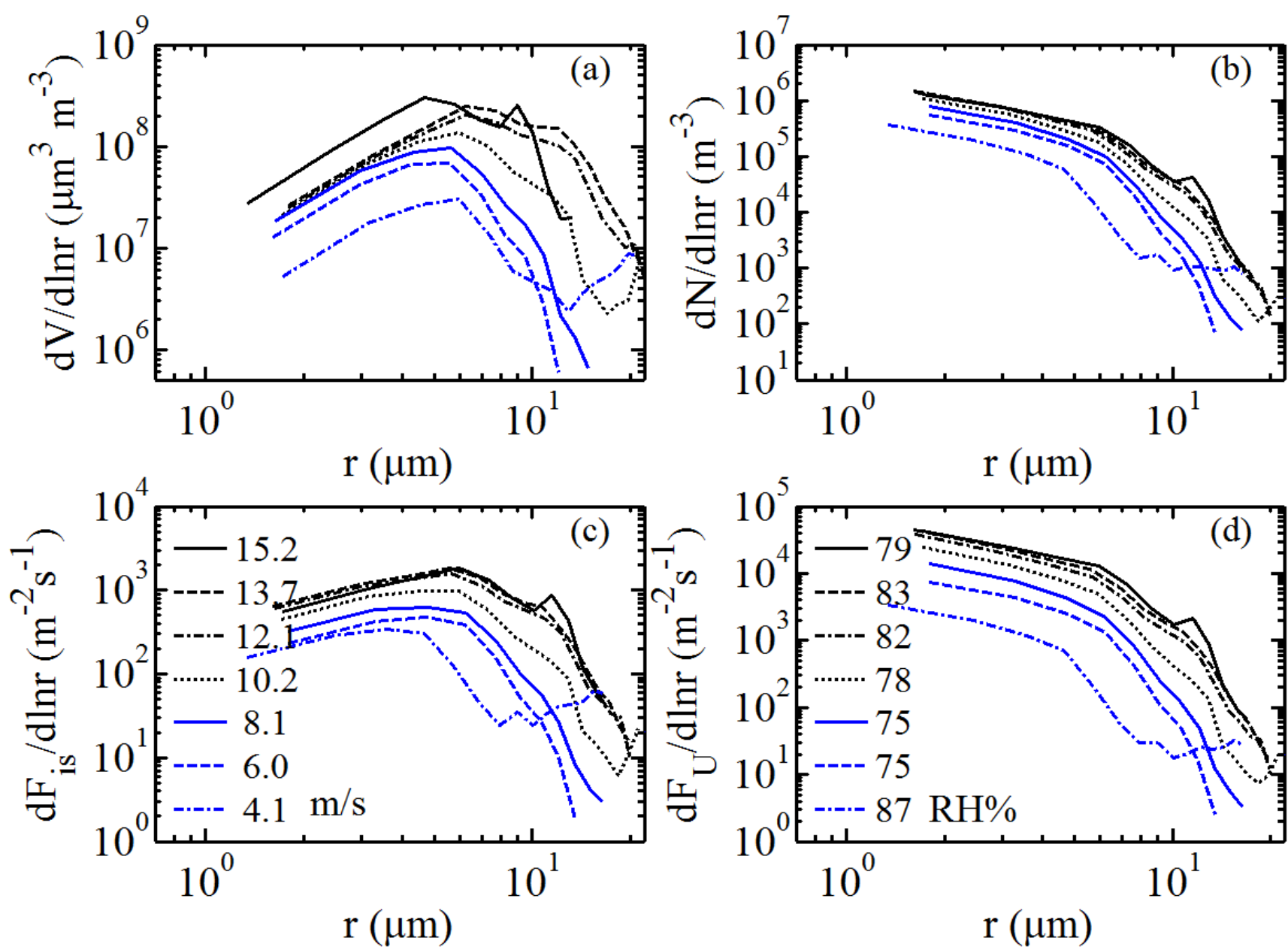

Fig. 7. The SSA size spectra: (a) Volume d$V$/dln$r$; (b) Number d$N$/dln$r$; (c) Flux computed with (3) and (4): d$F_{is}$/dln$r$; and (d) Flux computed with (5) and (6): d$F_U$/dln$r$.

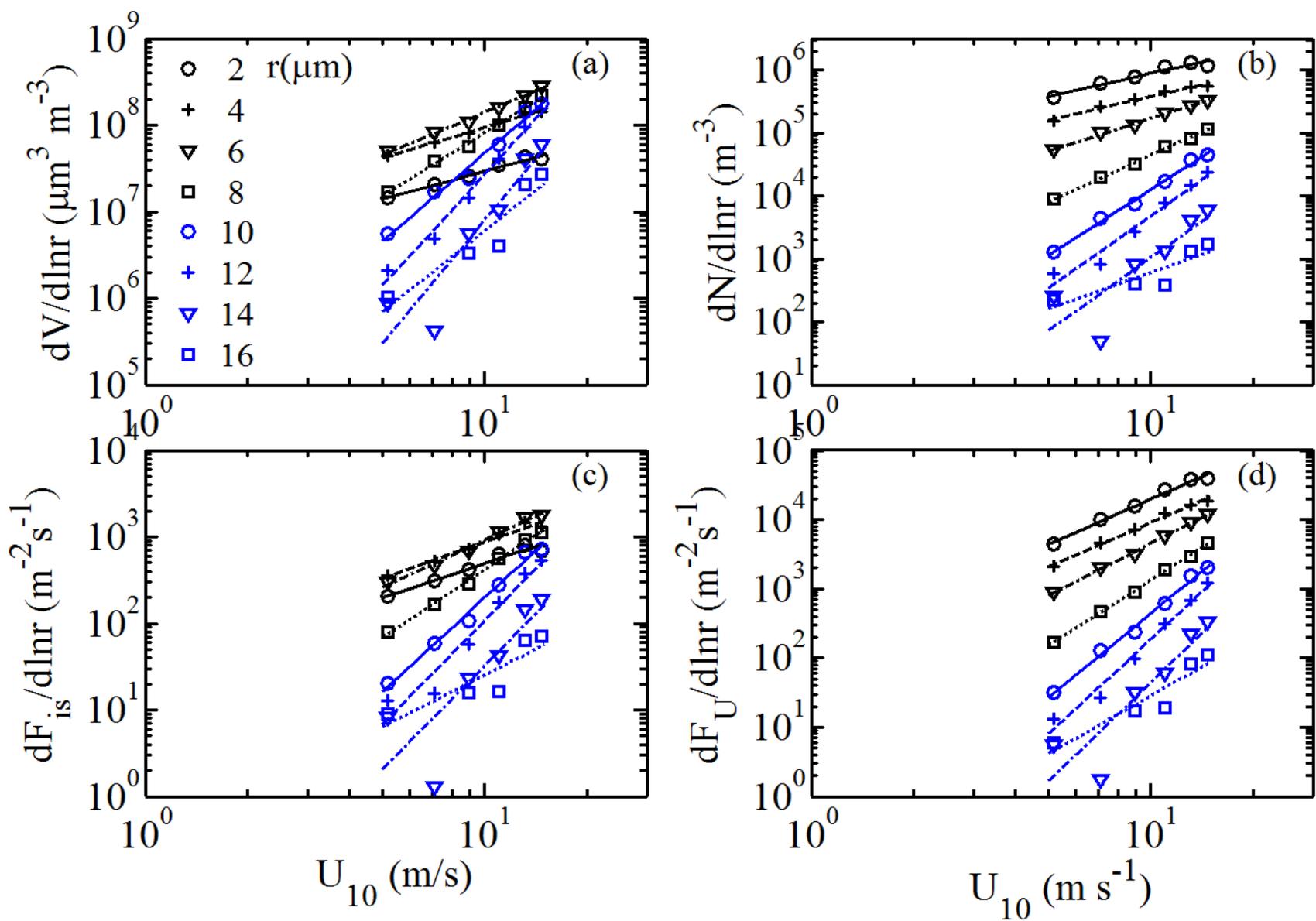

Fig. 8. Power-law relationship of the SSA size spectral components: (a) Volume d*V*/dln*r*; (b) Number d*N*/dln*r*; (c) Flux computed with (3) and (4): d*F_is*/dln*r*; and (d) Flux computed with (5) and (6): d*F_U*/dln*r*.

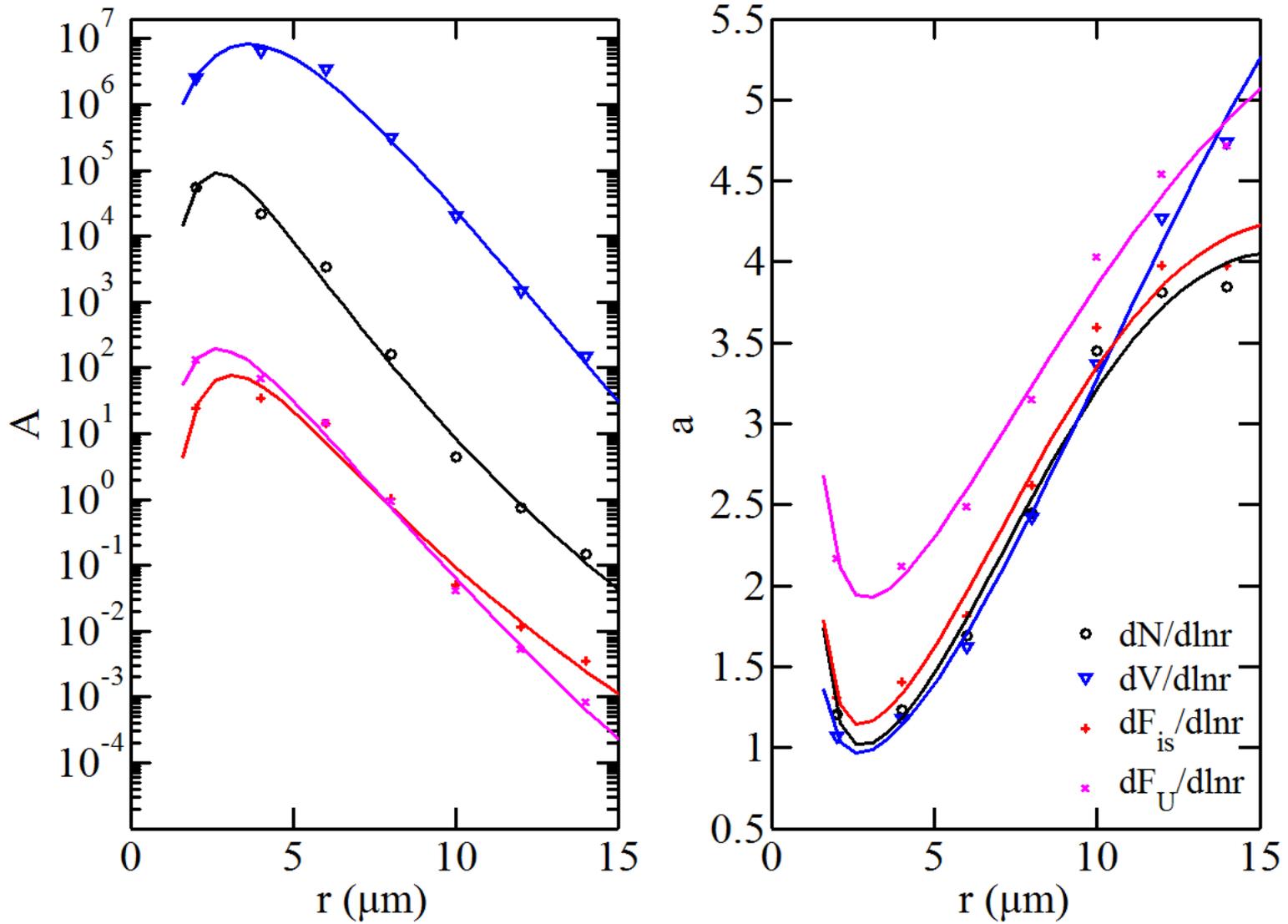

Fig. 9: The coefficients of power-law empirical model functions for the SSA size spectra, $dN/d\ln r$, $dV/d\ln r$, $dF_{is}/d\ln r$, and $dF_U/d\ln r$: (a) proportionality coefficient $A$, and (b) exponent $a$.

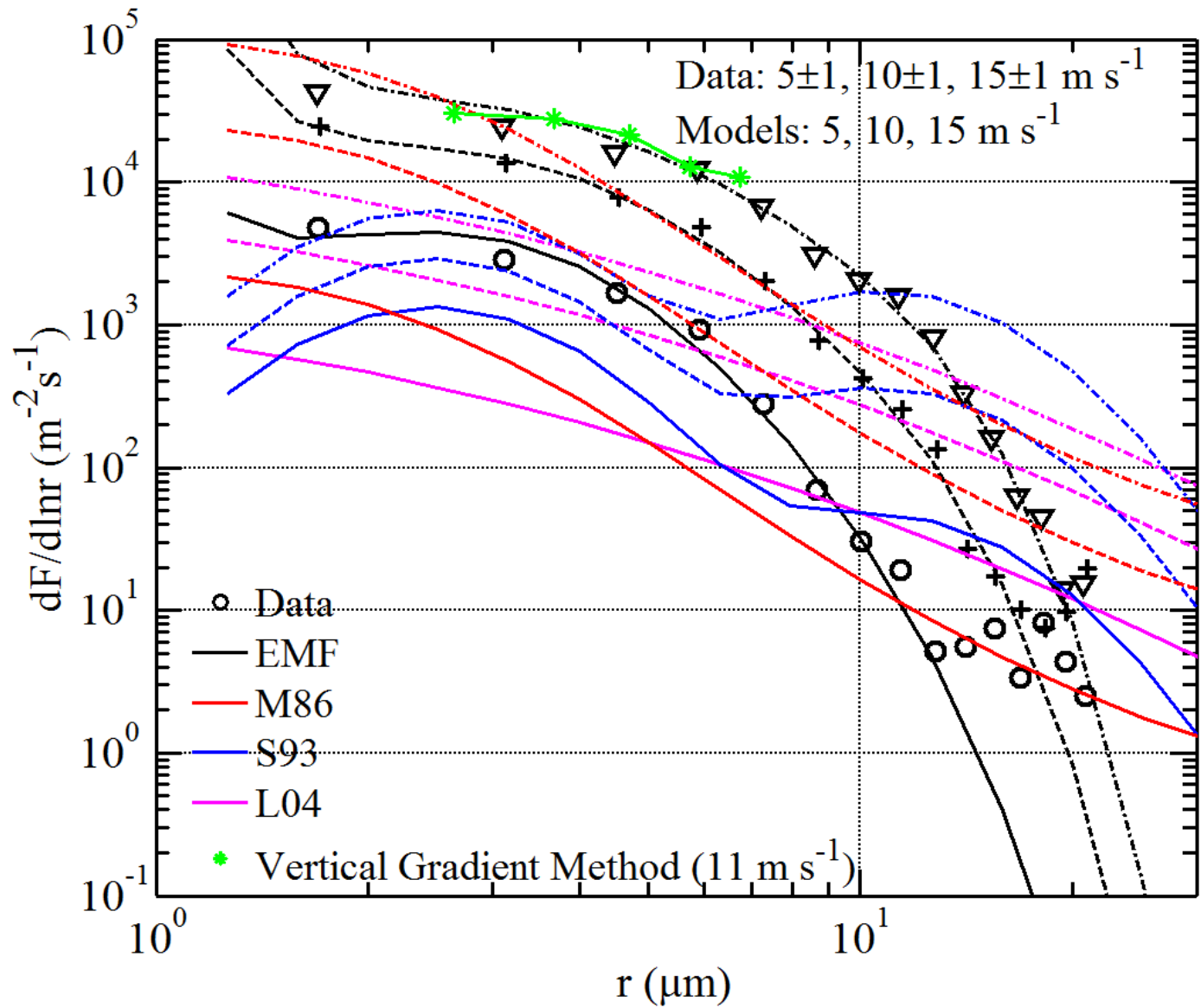

Fig. 10: Empirical model function (EMF) of $dF_U/d\ln r$, and its comparison with data and other model functions (Monahan et al. 1986: M86; Smith et al. 1993: S93; Lewis and Schwartz 2004: L04). Plotting symbols circle, plus and triangle show data at 5, 10 and 15 m s$^{-1}$, respectively; solid, dashed, and dashed-dotted curves show various models (described in the legend with different colors) at 5, 10 and 15 m s$^{-1}$, respectively.

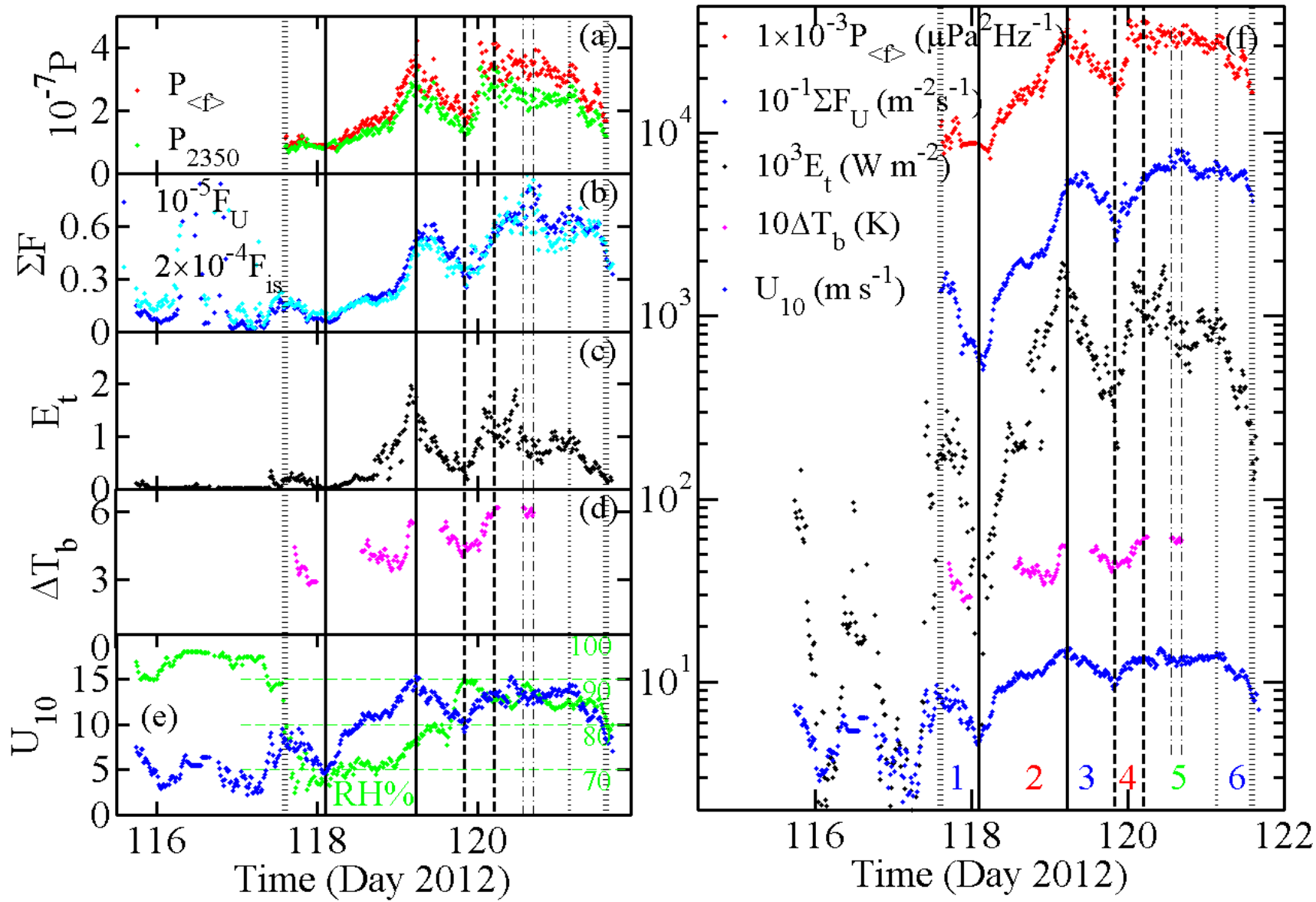

Fig. 11. Combined time series of underwater acoustic noise $P$, SSA flux $\Sigma F$, wind-sea energy dissipation rate $E_t$, roughness component of brightness temperature deviation $\Delta T_b$, and wind speed $U_{10}$. The time series are shown in linear scale in separate panels (a)-(e) on the left column; and together in logarithmic scale on the right column (f). Both $P_{2350}$ and $P_{<f>}$ are shown in (a); and both $\Sigma F_{is}$ and $\Sigma F_U$ are given in (b).

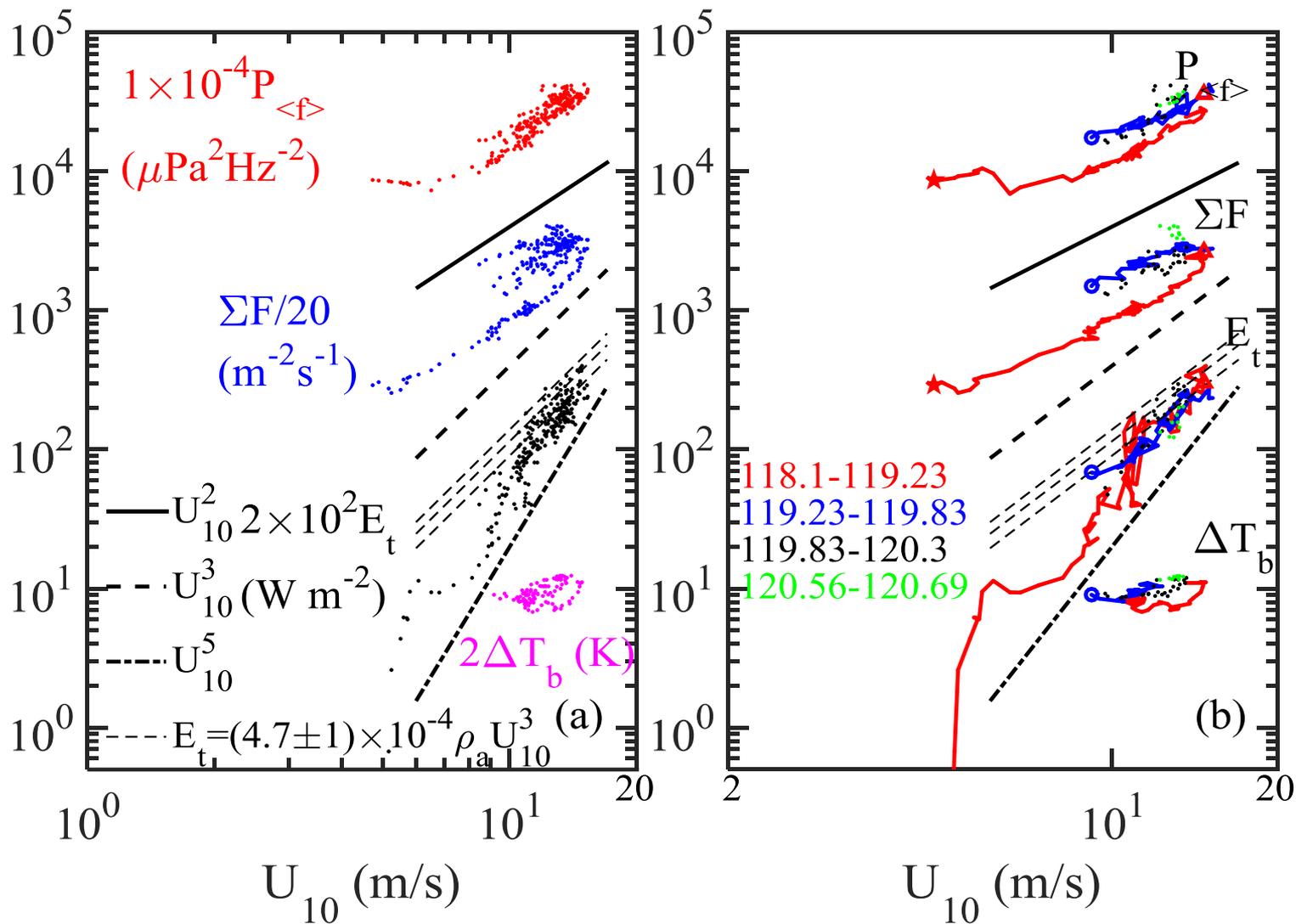

Fig. 12. Scatter plots showing the power-law dependence on $U_{10}$ for $P$, $\Sigma F$, $E_t$, and $\Delta T_b$: (a) Data unsorted; and (b) Sorting of 4 time segments illustrated in different colors. See text for more detail.

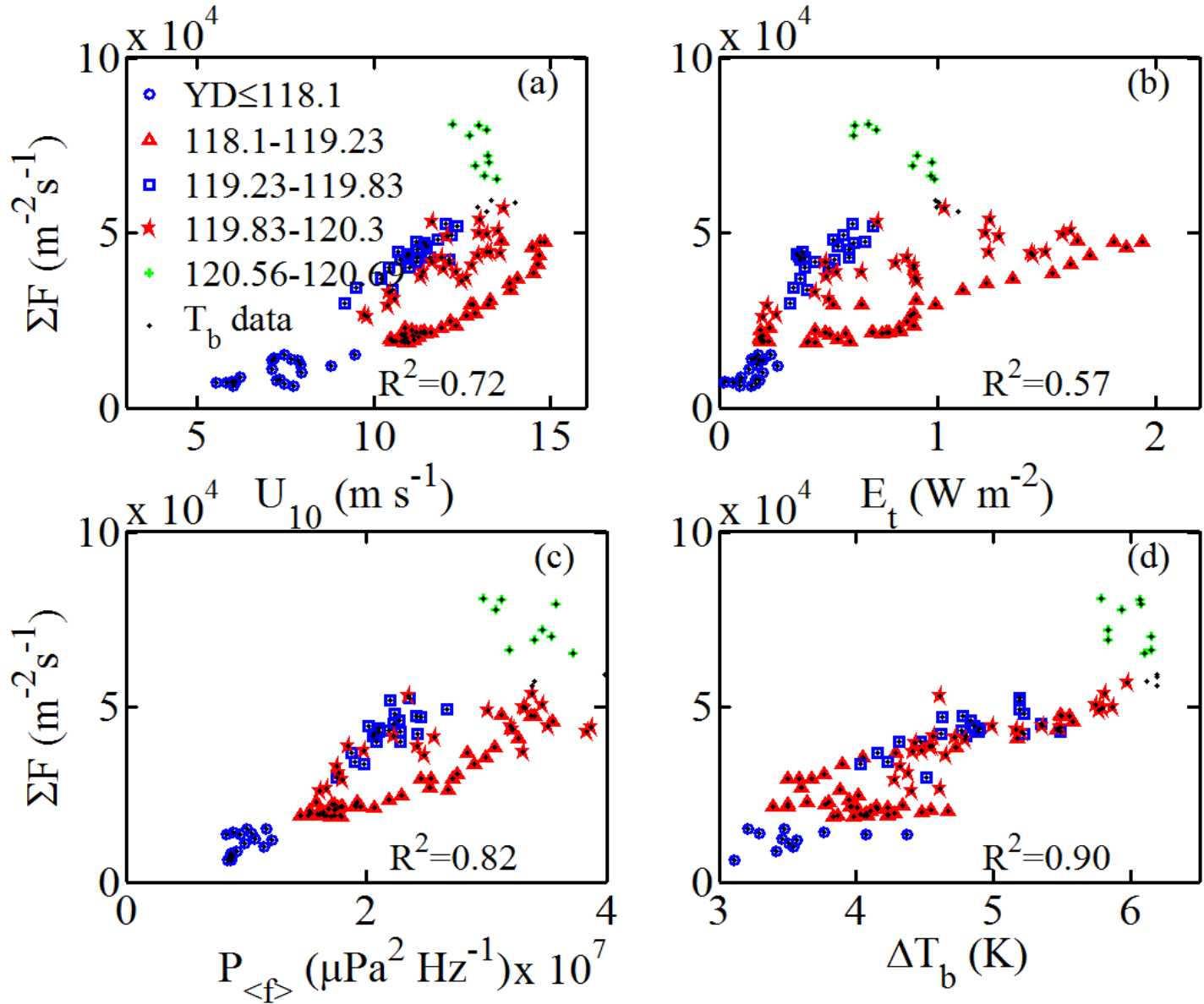

Fig. 13. Correlation between SSA flux $\Sigma F$ and (a) Wind speed $U_{10}$; (b) Wind-sea energy dissipation rate $E_t$; (c) Underwater acoustic noise $P_{<f>}$; and (d) Roughness component of brightness temperature deviation $\Delta T_b$; only co-registered data are shown (constrained mainly by $\Delta T_b$).

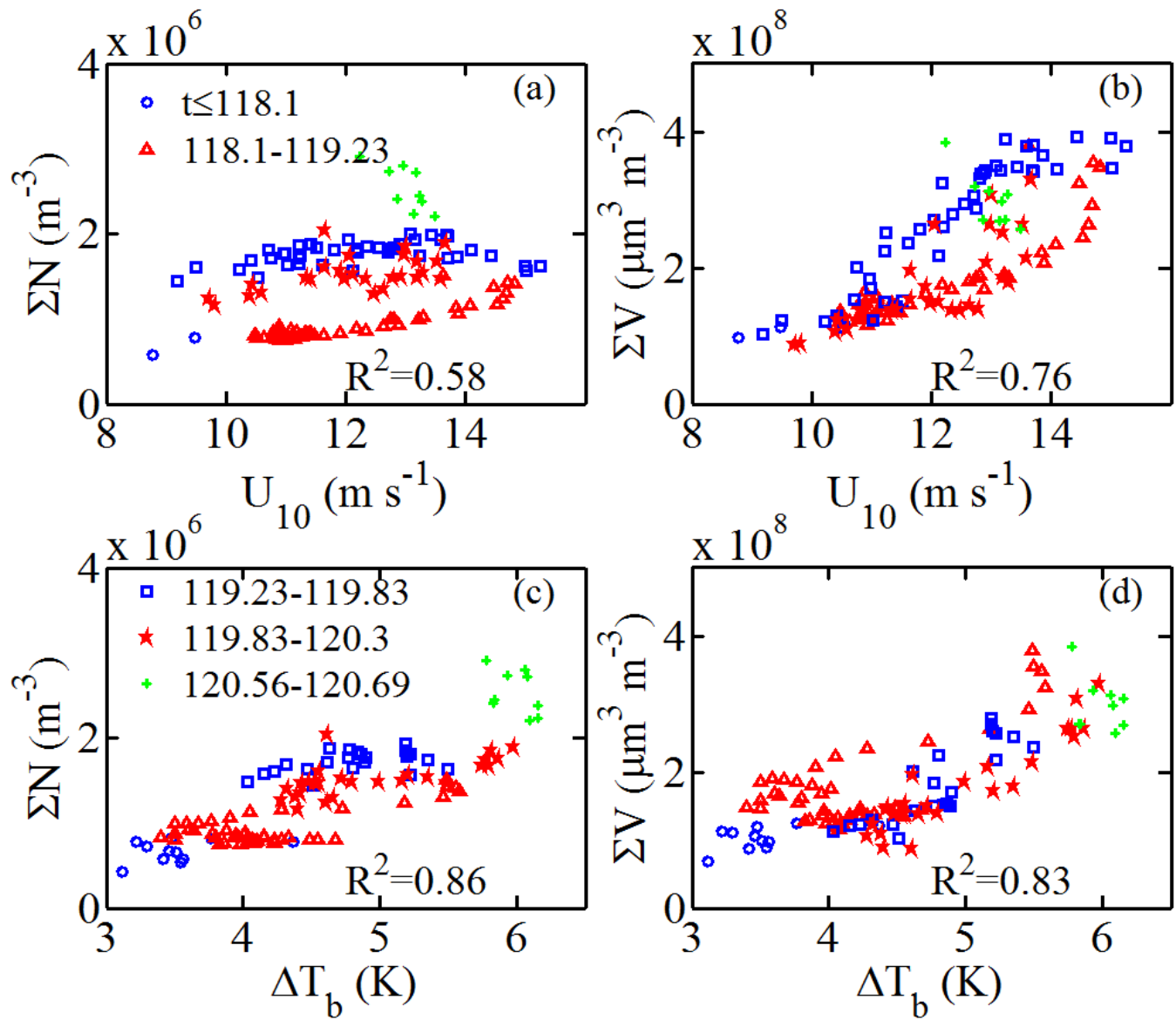

Fig. 14. Comparison of different SSA representations and their dependence on $U_{10}$ and $\Delta T_b$: (a) $\Sigma N(U_{10})$; (b) $\Sigma V(U_{10})$; (c) $\Sigma N(\Delta T_b)$; and (d) $\Sigma V(\Delta T_b)$; only co-registered data are shown (constrained mainly by $\Delta T_b$).

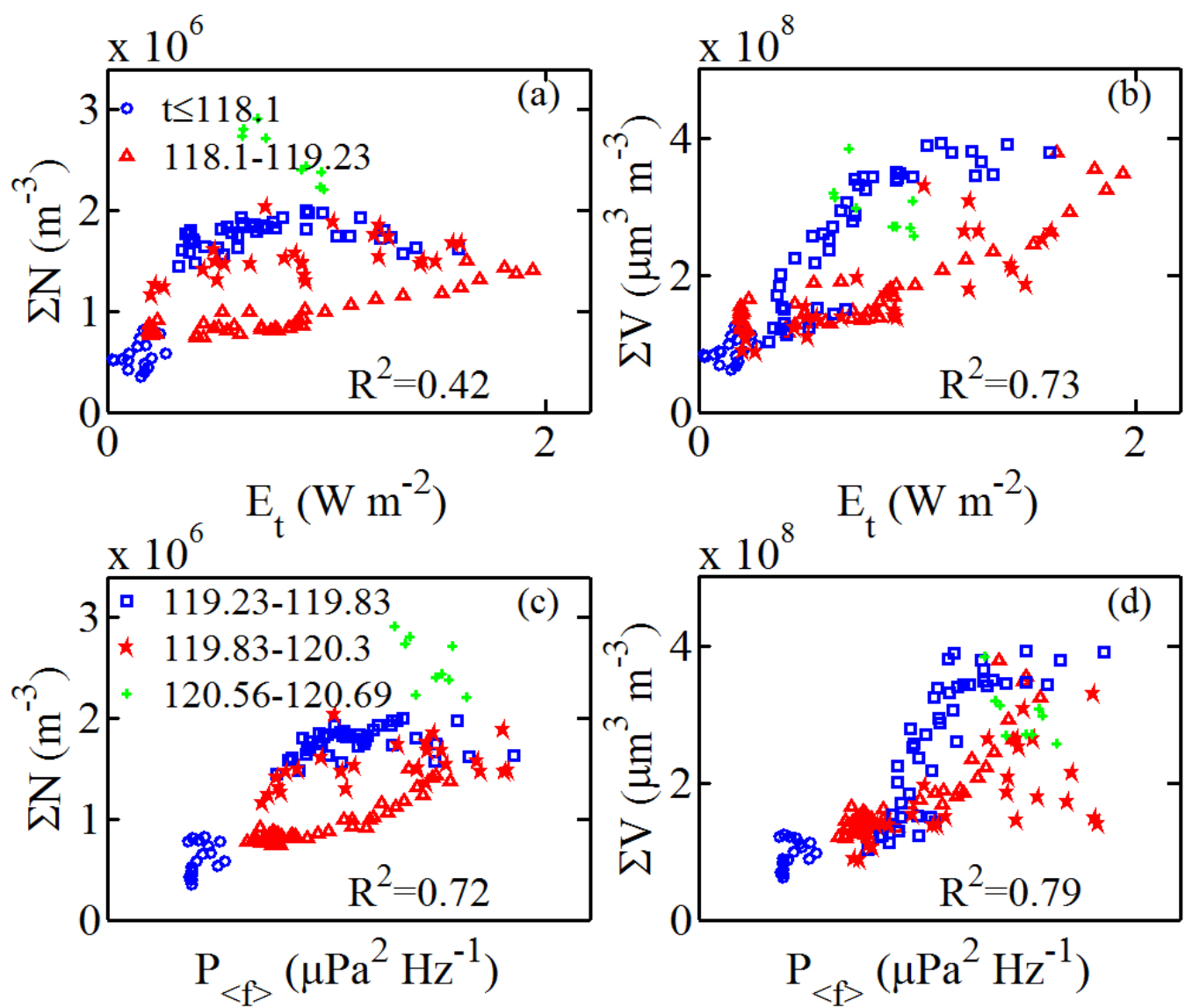

Fig. 15. Same as Fig. 14 but for dependence on $E_t$ and $P_{<f>}$.

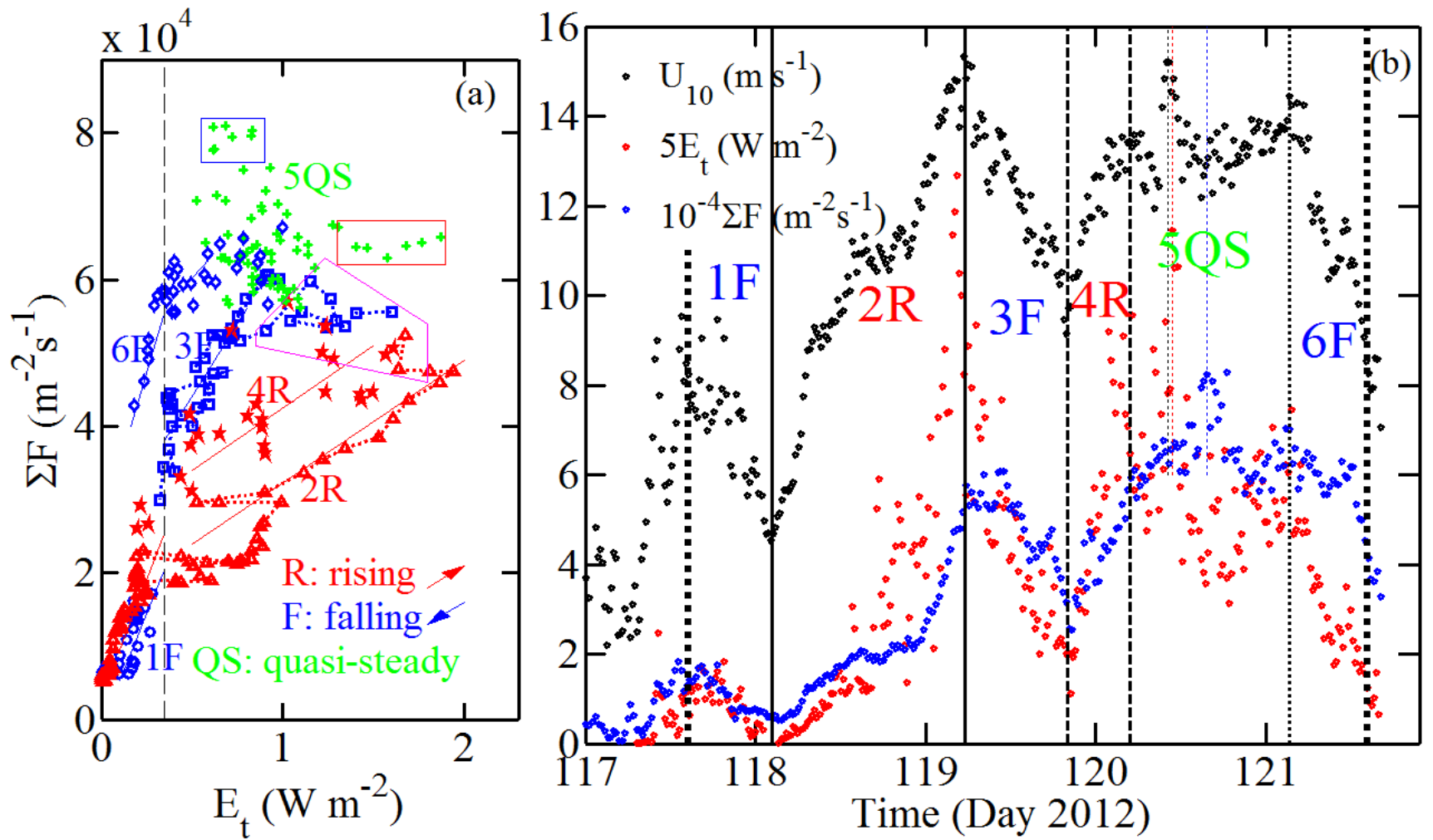

Fig. 16. (a) Episodic behavior of the SSA flux showing differences in the rate of change as a function of $E_t$ in rising wind, falling wind and quasi-steady conditions; and (b) Time series of $U_{10}$, $E_t$ and $\Sigma F$, with 6 episodes identified. The general trend of rising and falling winds in $\Sigma F(E_t)$ is indicated by the arrows in (a); effects of time lags between $U_{10}$, $E_t$ and $\Sigma F$ are illustrated with dotted lines connecting 2R and 3F events and the magenta box highlighting the transition segments, as well as the red and blue boxes of the 5QS event; see text for more discussion.

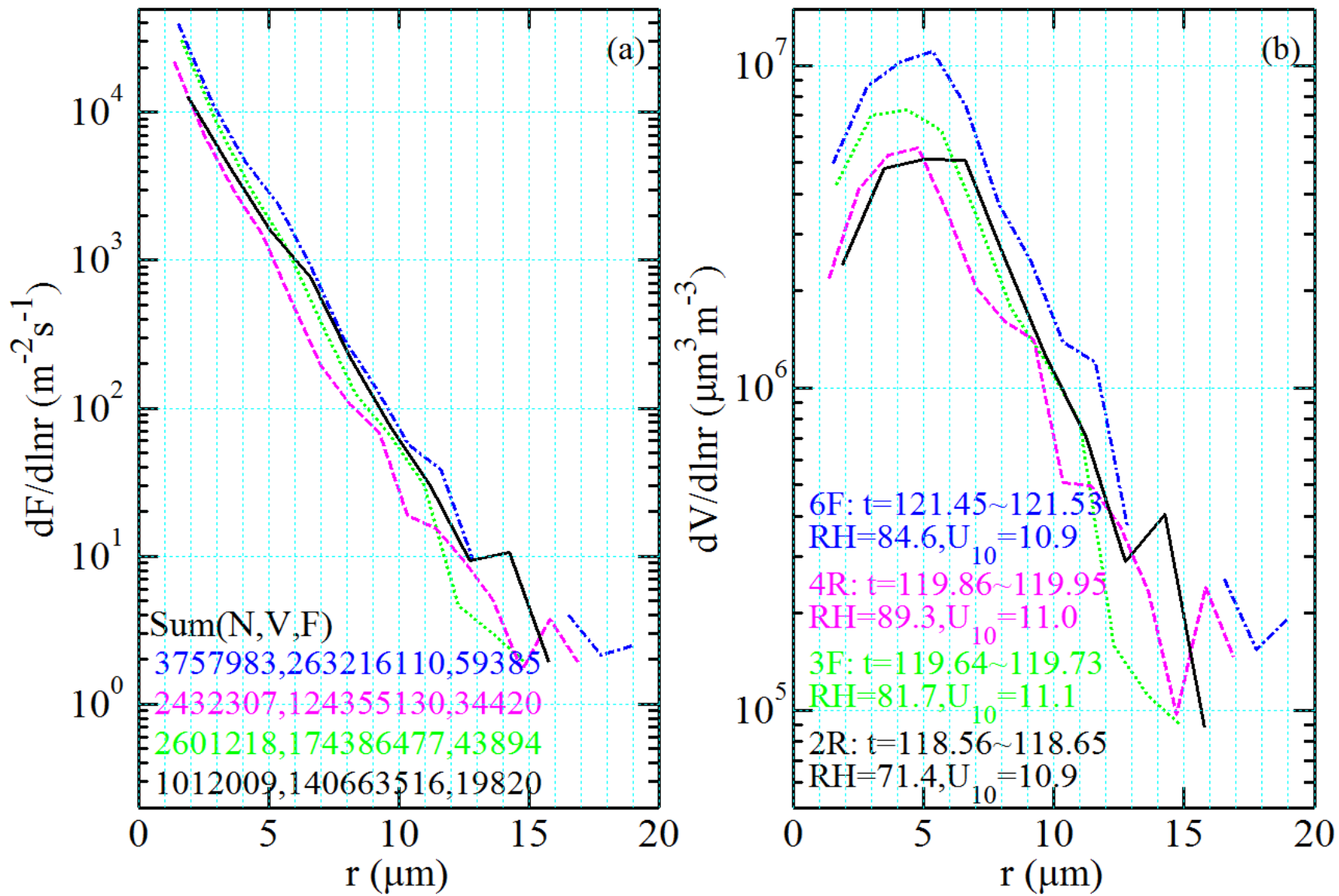

Fig. 17. Relative humidity effect on the SSA size spectra; here 2-h average spectra in 2 rising wind events and 2 falling wind events are illustrated, the average wind speeds are within $11.0 \pm 0.1$ m s$^{-1}$: (a) d$F$/dln$r$, and (b) d$V$/dln$r$.

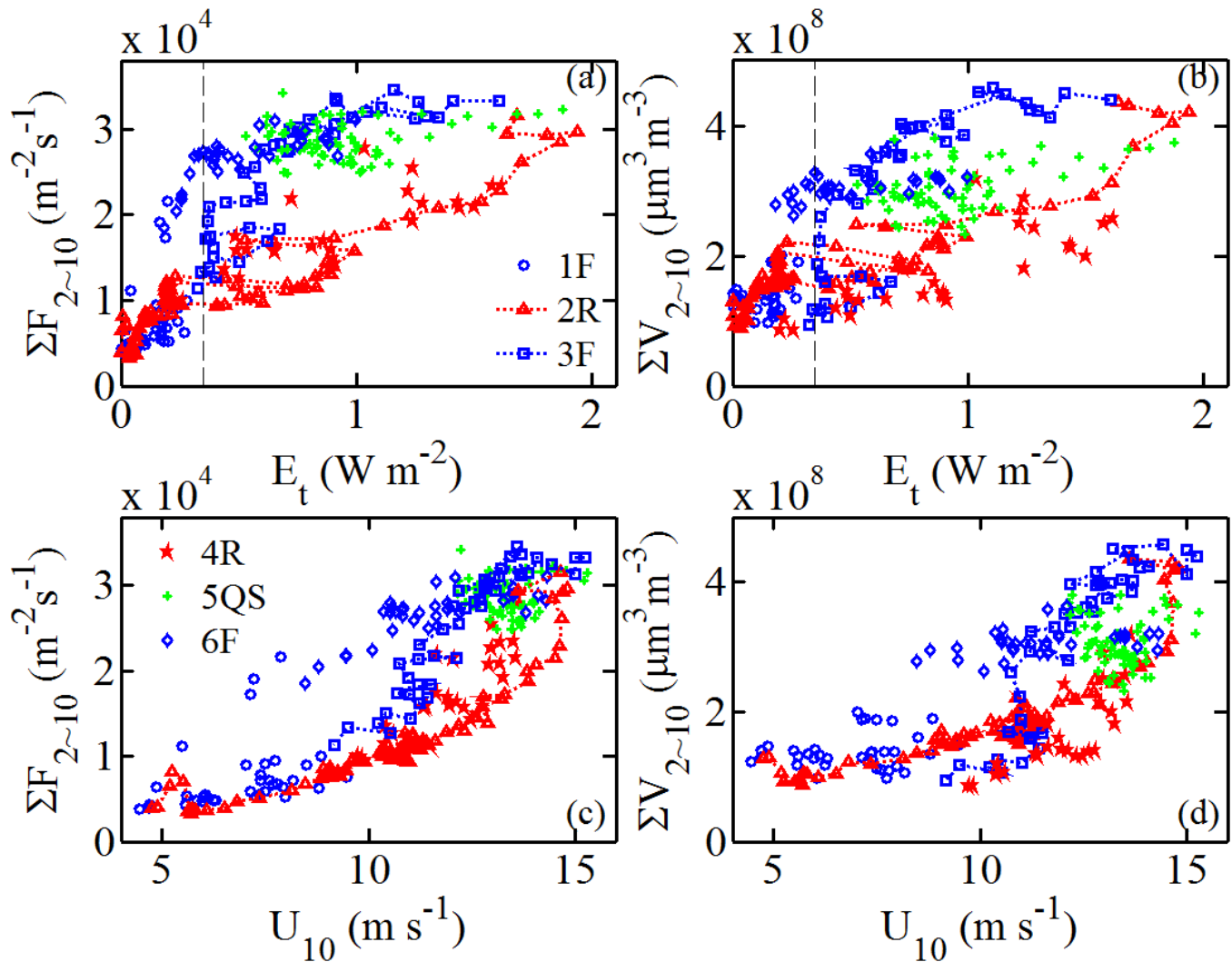

Fig. 18. Episodic behavior of the SSA production showing differences in the rate of change (in rising wind, falling wind and quasi-steady conditions) in terms of: (a) $\Sigma F_{2\sim10}(E_t)$; (b) $\Sigma V_{2\sim10}(E_t)$; (c) $\Sigma F_{2\sim10}(U_{10})$; and (d) (a) $\Sigma V_{2\sim10}(U_{10})$, where subscript 2~10 denotes integration over the size range $r_{80}=2$ to $10$ µm.

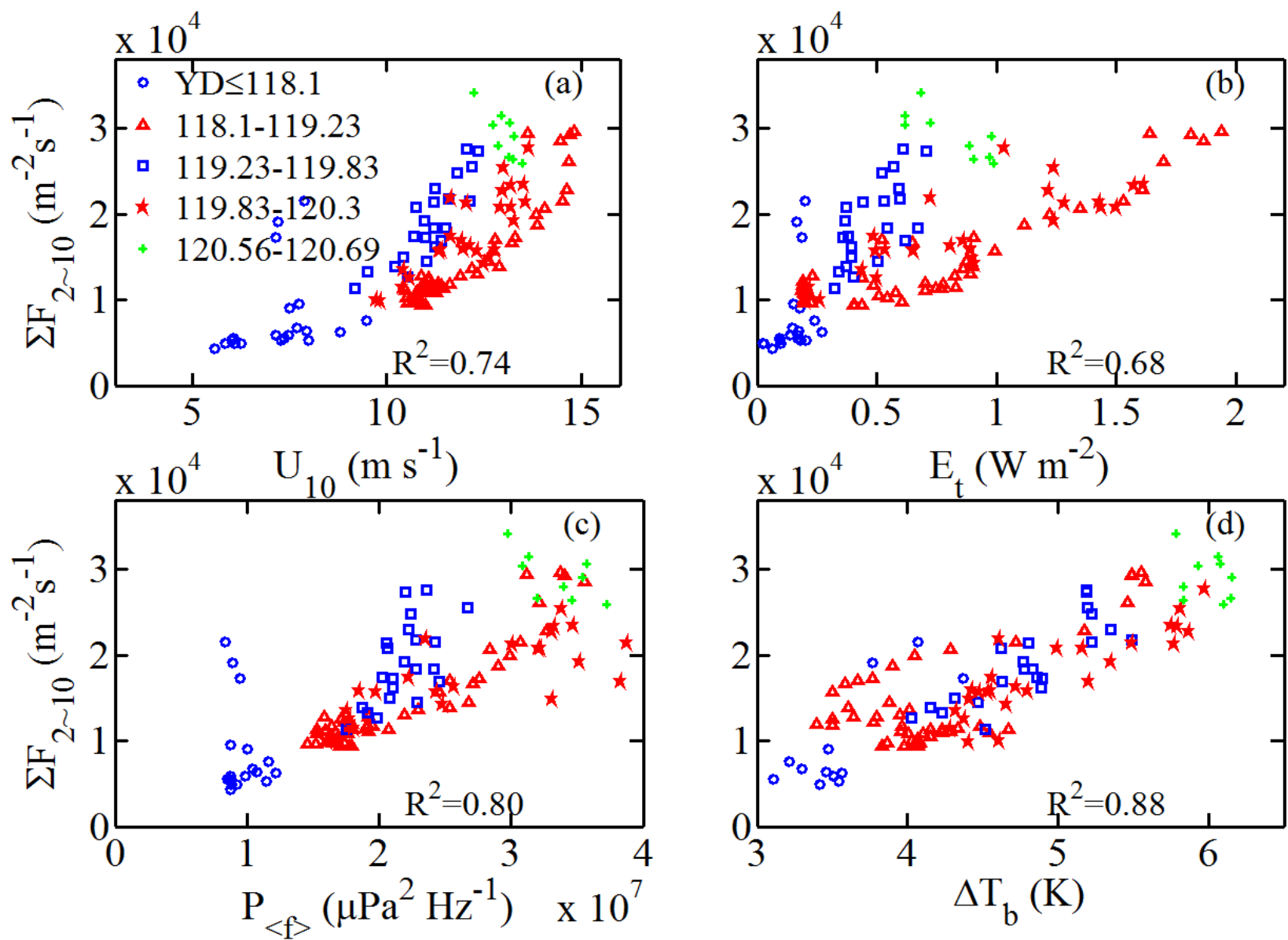

Fig. 19. Same as Fig. 13 but for integration over the size range $r_{80}$=2 to 10 µm.

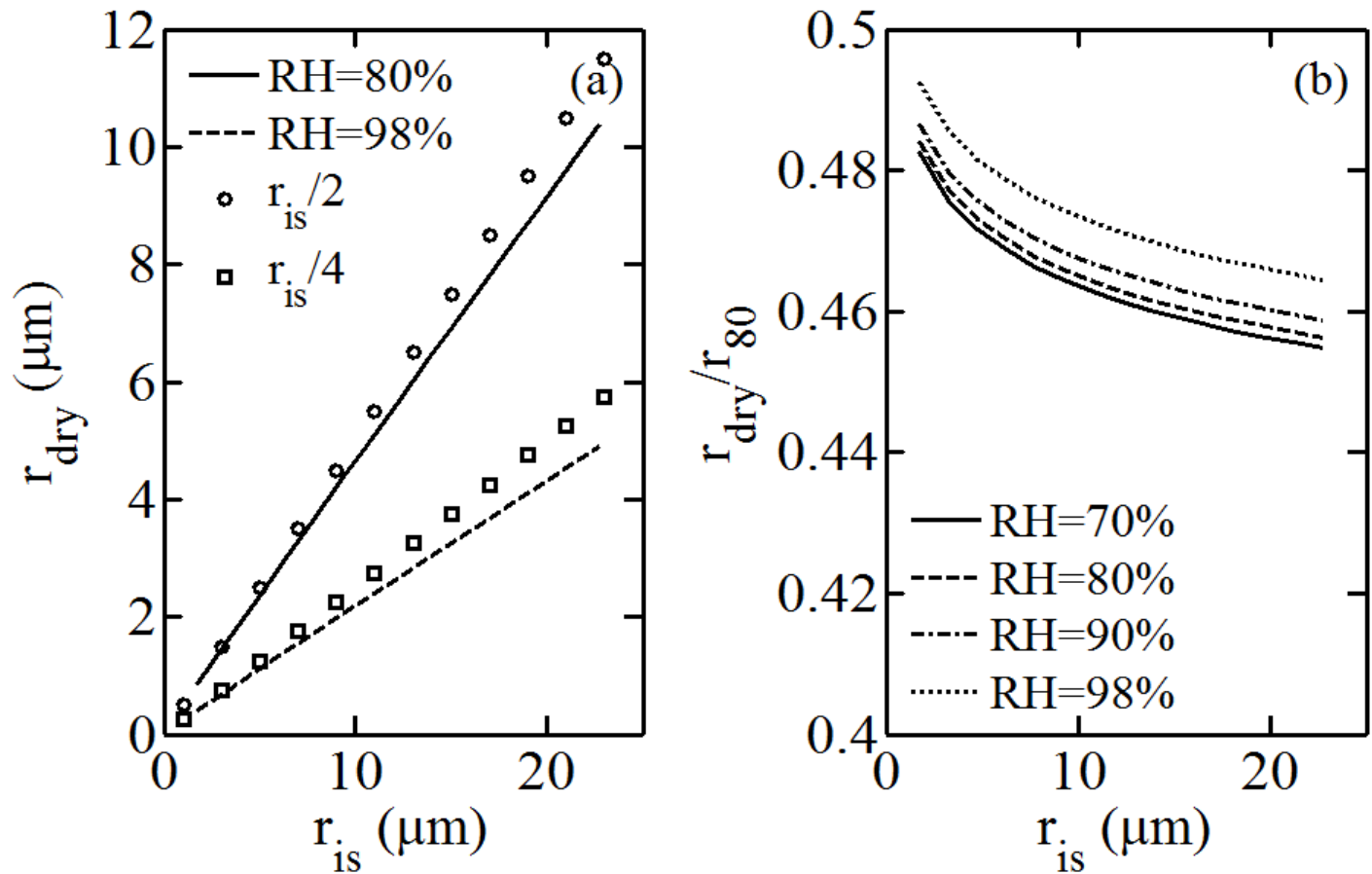

Fig. A1. (a) Comparison of two RH correction schemes of SSA radius used in this paper: curves (Gerber 1985) and symbols (Lewis and Schwartz 2004); (b) The ratio $r_{dry}r_{80}^{-1}$ as a function of RH based on Gerber (1985).